\documentclass[12pt,preprint2]{aastex}

\usepackage{lscape}

\shorttitle{Disks in the Arches cluster}
\shortauthors{A. Stolte et al.}

\begin{document}

\title{Disks in the Arches cluster -- survival in a starburst environment\altaffilmark{*}}
\author{A. Stolte\altaffilmark{1,2}, M. R. Morris\altaffilmark{2}, A. M. Ghez\altaffilmark{2,3}, T. Do\altaffilmark{2}, J. R. Lu\altaffilmark{4}, S. A. Wright\altaffilmark{5}, C. Ballard\altaffilmark{6}, E. Mills\altaffilmark{2},  K. Matthews\altaffilmark{4}}
\altaffiltext{*}{Based on observations made with the Keck II telescope on Mauna Kea, Hawai'i,
and with ESO Telescopes at the Paranal Observatory under programme ID 60.A-9026.}
\altaffiltext{1}{I. Physikalisches Institut, Universit\"at zu K\"oln, Z\"ulpicher Str. 77, 50937 K\"oln, Germany, astolte@ph1.uni-koeln.de}
\altaffiltext{2}{Division of Astronomy and Astrophysics, UCLA, Los Angeles, CA 90095-1547, ghez@astro.ucla.edu, morris@astro.ucla.edu, jlu@astro.ucla.edu}
\altaffiltext{3}{Institute of Geophysics and Planetary Physics, UCLA, Los Angeles, CA 90095} 
\altaffiltext{4}{Caltech Optical Observatories, California Institute of Technology, MS 320-47, Pasadena, CA 91225, kym@caltech.edu}
\altaffiltext{5}{UC Berkeley, Astronomy Department, 601 Campbell Hall, Berkeley, CA 94720-3411, saw@astro.berkeley.edu}
\altaffiltext{6}{14090 Robler Road, Sherman Oaks, CA 91423}

\begin{abstract}
Deep Keck/NIRC2 $HK'L'$ observations of the Arches cluster near the Galactic 
center reveal a significant population of near-infrared excess sources.
We combine the $L'$-band excess observations with $K'$-band proper motions, 
which allow us to confirm cluster membership of excess
sources in a starburst cluster for the first time. The robust removal
of field contamination provides a reliable disk fraction down to 
our completeness limit of $H=19$ mag, or $\sim 5\,M_\odot$ 
at the distance of the Arches.
Of the 24 identified sources with $K'-L' > 2.0$ mag, 21 have reliable 
proper motion measurements, all of which are proper motion members of 
the Arches cluster. VLT/SINFONI $K'$-band 
spectroscopy of three excess sources reveals strong CO bandhead emission,
which we interpret as the signature of dense circumstellar disks.
The detection of strong disk emission from the Arches stars is surprising 
in view of the high mass of the B-type main sequence host stars of the disks
and the intense starburst environment. We find a disk fraction of 
$6 \pm 2$\% among B-type stars in the Arches cluster. A radial increase in 
the disk fraction from 3 to 10\% suggests rapid disk destruction in the 
immediate vicinity of numerous O-type stars in the cluster core.
A comparison between the Arches and other high- and low-mass star-forming 
regions provides strong indication that disk depletion is significantly 
more rapid in compact starburst clusters than in moderate star-forming 
environments.
\end{abstract}


\section{Introduction}
\label{intro}

\subsection{The Arches starburst cluster}

The Arches cluster is a dense, massive, young stellar cluster
located in the central molecular zone near the center of the Galaxy
(Cotera et al.~1996, Figer et al.~1999, 2002, Stolte et al.~2002, 2005, 
Kim et al.~2006, Stolte et al.~2008, Espinoza et al.~2009).
With a total stellar mass of $2\times 10^4\,M_\odot$ (Espinoza et al.~2009), 
it is considered 
one of the rare starburst clusters in the Milky Way. The young age
of only $2.5 \pm 0.5$ Myr (Najarro et al.~2004) places it at an
evolutionary state comparable to nearby star-forming regions.
Nearby star-forming regions at these young ages, as observed, for example, 
in the Orion complex, display a significant
fraction of circumstellar disks remaining from the star formation process
(Haisch et al.~2001, Hern\'{a}ndez et al.~2007, and references therein).
As a starburst cluster, however, the Arches hosts 125 O-type stars
(Figer et al.~2002, Stolte et al.~2005), 
some of which have already evolved to the earliest Wolf-Rayet phases 
(Martins et al.~2008). For comparison, the Orion nebula cluster hosts 
only 2 to 4 O-type stars (Hillenbrand 1997). 
Most of our knowledge of the first few 
million years of stellar evolution and disk depletion 
stems from nearby star-forming regions of significantly lower density 
and stellar mass, while the effects of the starburst environment on circumstellar
material and the disk lifetime has rarely been probed. 

Increasing evidence for remnant accretion disks around young, massive
OB stars is deduced from K-band spectroscopy in very young star-forming
regions (e.g., M17: Hanson et al.~1997, Hoffmeister et al.~2006, 
NGC 3576: Blum et al.~2004, Figueredo et al.~2005)
and UCH{\sc II} regions (e.g., Bik et al.~2005, 2006, and references
therein). The large fraction of near-infrared excess objects
and the fact that early O-type stars are on the main sequence
suggest very young ages of less than 1-2 Myr for these environments.
Observing disk fractions at later evolutionary stages is complicated
by the steeply decreasing disk fraction as clusters age (Haisch et al.~2001), 
and the 
decreasing brightness of the increasingly depleted disks themselves.
The Arches cluster, at an age of $2.5 \pm 0.5$ Myr (Najarro et al.~2004),
fills one of the rare gaps in cluster evolution. Not only does it 
provide an estimate of the disk lifetime in its substantial B-star
population, it additionally connects disk survival with the dense 
environment of a starburst cluster.

\subsection{Circumstellar disks in young star clusters}

In young star-forming regions with ages less than 10 Myr, $L$-band
excess provides one of the most efficient tools to detect circumstellar 
disks (Haisch et al.~2001, see also Hillenbrand et al.~1992, Lada et al.~2000). 
As the illumination
of the dense, inner disk rim by the central star generates 3.8$\mu$m
emission from hot dust (Natta et al.~2001), 
the excess $L$-band flux indicates dense disks surviving from the 
star-formation process. The hostile starburst cluster environment 
can accelerate disk destruction, to the extent that one might not 
expect to find any remaining gaseous disks in a cluster like the Arches.
The finding of disks provides crucial clues for our understanding of disk 
survival and the likelyhood of planet formation in dense cluster environments.

A linear decrease is observed in the disk fraction as a function of  
increasing cluster age in nearby star-forming regions 
(Haisch et al.~2001). This decrease in near-infrared
excess emission can be understood as the depletion of dust
in the inner, hot disk rim due to photoevaporation and grain growth
(e.g., Takeuchi et al.~2005). The lifetimes of young disks, as derived from 
near-infrared excess fractions, indicate that 
most disks are depleted within 10 Myr, with shorter disk lifetimes 
for higher mass stars. In a pioneering study of 47 Herbig Ae/Be stars,
Hillenbrand et al.~(1992) found a more rapid depletion of disks around
B-type stars with masses of $\sim 3-30\,M_\odot$ than in A stars in
the $1-3\,M_\odot$ mass range. On average, the disks around high-mass stars 
do not survive for more than 1 Myr, with a characteristic disk age 
of only a few $10^5$ yr (Alonso-Albi et al.~2009).

According to disk depletion simulations, the timescale for disk survival 
is expected to depend on the mass of the central star and 
on the ambient UV radiation field, and thus on the cluster environment.
However, disk photoevaporation models do not arrive at a consistent conclusion 
for disk survival times around high-mass stars.
Models of disks around high-mass O- and early B-type stars in isolation
suggest that while the outer disk is evaporated by the central
star on timescales of less than 1 Myr, the inner, dense disks with
radii of $< 10$ AU can survive
for several Myr in late O-type stars, and more than 20 Myr in early 
B-type stars (Richling \& Yorke 1997, Hollenbach et al.~2000, 
Scally \& Clarke 2001, see also Bik et al.~2006). 
Photoevaporation models by Takeuchi et al.~(2005), on the other hand, 
predict that the dense, inner disk around a Herbig Ae/Be 
star will be depleted out to a radius of 40 AU within 3 Myr. In a starburst cluster,
the numerous O- and B-type stars additionally create a strong EUV/FUV radiation 
field that accelerates disk evaporation. Simulations by Armitage (2000) suggest 
that a disk with a dust mass of $M_{d} = 0.04\,M_\odot$ can survive for up to 1 Myr 
in clusters with $10^4$ stars at a distance of 1 pc from 
the dense core, while the same disk will be evaporated in less than $10^5$ yrs if 
located at 0.1-0.3 pc from the massive stars in the cluster center.
Similar conclusions are reached by Fatuzzo \& Adams (2008), who simulate the 
influence of the radiation field on circumstellar disks in a massive star cluster 
with respect to planetary system formation.
Tidal forces during dynamical interactions
between cluster members might additionally truncate circumstellar disks
(Olczak et al.~2008, Pfalzner et al.~2006, Cesaroni et al.~2007, 
Scally \& Clarke 2001, Boffin et al.~1998), and enhance the exposure of the inner 
disks to ambient UV radiation, thereby accelerating dust depletion. 

In the Arches, NGC\,3603, and similarly dense starburst clusters, the situation 
may be more extreme than in nearby star-forming regions, as the numerous O-type 
stars segregate toward the cluster core,
where the combined effect of their radiation fields act on circumstellar material 
around lower-mass cluster members as well. The radiation environment in starburst 
clusters therefore has important consequences for the survival timescale of disks,
both in the Milky Way and in extragalactic systems, and thus
for the potential of planetary system formation in dense star-forming environments.
In particular, the disk fraction is predicted to be a function of the 
stellar density in addition to the cluster age. A lower disk fraction in 
high-density environments might influence the accretion timescale and final
stellar masses, and hence the initial stellar mass function.
Observationally, the effect of the ambient stellar density can be evidenced 
as a depletion of 
circumstellar disks towards the dense cluster core, where the most massive
stars reside. A decrease in the disk fraction with decreasing cluster radius
is observed in the young starburst cluster at the center of the giant H{\sc II}
region NGC\,3603 (Harayama et al.~2008, Stolte et al.~2004).
As a consequence of limited spatial resolution and sensitivity in $L$-band
observations, the high foreground extinction and large distances to Milky Way
starburst clusters has so far hampered a systematic study of disk survival in the 
starburst environment.
Thus far, NGC\.3603\,YC is the only starburst cluster where the disk fraction
has been determined.

Fortunately, our study of the Arches cluster has provided an opportunity 
to progress in this domain. 
We have obtained $HK'L'$ observations of the Arches cluster with the Keck II
laser-assisted adaptive optics system (LGS-AO) during our proper motion 
campaign to study the kinematics, extinction and stellar mass distribution
of the cluster core. In the central $r < 0.5$ pc of the cluster 
core, we detect 24 sources with substantial 3.8$\mu$m excesses of $\Delta(K'-L') \ge 0.7$ mag
as compared to the cluster main sequence population. VLT/SINFONI K-band spectroscopy
of three of these excess sources reveals CO emission, indicative of disk emission. 
The high contamination 
of field stars in the inner Galaxy particularly influences the fraction of main 
sequence sources at the faint end of the cluster main sequence population.
The combination of a complete proper motion member sample among both the 
disk-bearing $L$-band excess sources and main sequence stars allows us to 
obtain the first measurement of the disk fraction in the Arches cluster.
With the aim of assessing the prediction of an environment-dependent disk lifetime, 
we determine the disk fraction as a function of radius and discuss the prospects 
for disk survival around high-mass stars in a dense cluster environment.

The observations of the Arches cluster are presented in Sec.~\ref{obssec}.
Results are presented in Sec.~\ref{excsec}, where 
the $K'-L'$ excess sample is defined in Sec.~\ref{diskexcsec}, K-band spectroscopy
of selected sources is analysed in Sec.~\ref{diskspecsec}, and the
disk fraction of the Arches is derived in Sec.~\ref{diskfracsec}.
The results are discussed in Sec.~\ref{discsec}, including a comparison to 
similarly young star-forming regions (Sec.~\ref{compsec}), 
and our results are summarised in Sec.~\ref{sumsec}.

\section{Observations}
\label{obssec}

\subsection{Keck/NIRC2 $HK'L'$ imaging}
\label{photsec}

\subsubsection{Data reduction and PSF extraction}

Five fields in the Arches cluster were observed in $HK'L'$ with the Keck II
NIRC2 camera (PI: K.~Matthews) with the laser guide star adaptive optics system
(LGS-AO; Wizinowich et al.~2006, van Dam et al.~2006) between 2006 May and 2008 July.
The field and tip-tilt star positions, distance from the brightest source 
in the cluster core and to the tip-tilt source are provided in Table~\ref{fieldtab}.
The central wavelengths and bandwidths for these NIRC2 filters are 
$\lambda_H = 1.633\mu$m, $\delta\lambda_H = 0.296\mu$m, 
$\lambda_{K'} = 2.124\mu$m, $\delta\lambda_{K'} = 0.351\mu$m, 
$\lambda_{L'} = 3.776\mu$m, and $\delta\lambda_{L'} = 0.700\mu$m.
The narrow camera setting with a pixel scale of 
$0.00996^{''} {\rm pix^{-1}}$ delivered a field of view (FOV) of 
$10^{''} \times 10^{''}$ (Ghez et al.~2008). The LGS-AO system 
provided an $R=10$ mag wavefront reference source centered in the field of view. 
Two fainter natural guide stars with $R=15.3$ mag were used as tip-tilt 
reference sources, located $10^{''}-15^{''}$ from the center 
of each field. The adaptive optics performance was strongly dependent 
on the observing wavelength and on nightly atmospheric conditions.
The average Strehl ratios were 0.15 in $H$, 0.27 in $K'$, and 0.64 in $L'$.
A $K'$ mosaic of the five $HK'L'$ fields is illustrated in Fig.~\ref{mosaic},
and field positions and tip-tilt star offsets are provided in Table~\ref{fieldtab}.
A summary of the observations, including the observational details, 
the number of frames entering each final combined image, as well as
the resulting Strehl ratios and spatial resolutions, can be found 
in Table~\ref{obstab}.

The images were reduced with our NIRC2 python/pyraf pipeline (Lu et al.~2009),
which includes dark and flatfield correction, sky subtraction,
the removal of NIRC2 bad pixels and cosmic rays, and 
distortion correction (Thompson et al.~2001) prior to image combination. 
Ten separate sky frames were observed in each filter at the 
end of each 3 hour Arches observing block. From these open-loop,
star-free exposures, a median sky frame was created for each filter,
which was scaled to the median background value of each individual 
exposure prior to subtraction.
In $L'$, a special procedure was employed to additionally account for 
structured thermal background caused by the Keck II field rotator mirror. 
Residual background structure was minimized by observing sky frames
at the same rotator mirror angle as the science observations. 
Skies were taken with rotator angle increments of two degrees, and 
median skies were averaged from three adjacent rotator angle positions. 
We also note that the best performance is achieved 
in $L'$ when sky frames are taken close in time to the observations.
When sky frames were observed within one hour of our $L'$ targets,
the photometric sensitivity in the reduced images was typically 0.5 mag
deeper than on the images observed earlier during the same night. 
Strehl ratios and full-width at half maximum (FWHM) values of the 
point-spread function (PSF) 
were derived for each individual image during the pipeline reduction.
In observing sequences with less than 20 frames, all frames with closed-loop
AO correction were combined into the final image.
In observing sequences with more than 20 frames, frame selection to 
enhance the spatial resolution was possible without compromising 
photometric depth. In these sequences, only frames with 
$FWHM < 1.25 \times FWHM_{min}$ were included in the 
final image, where $FWHM_{min}$ was the minimum $FWHM$ achieved in the 
sequence for each field and filter. This rejection criterion typically 
removed between 10\% of the images under stable atmospheric conditions 
and 30\% of the images in the case of variable seeing and AO performance 
prior to image combination.
When frame selection was applied, the selected images were weighted 
by the Strehl ratio to enhance the resolution on the combined image.
After the standard reduction procedures, the drizzle algorithm was 
used to combine the images to preserve the original resolution and to
achieve the maximum sensitivity on the co-added image (Fruchter \& Hook 2002).

Relative photometry was extracted using the starfinder algorithm (Diolaiti et al.~2000),
which is designed to fit an empirical PSF to crowded stellar fields. 
The code has the advantage of creating an empirical 2-dimensional 
PSF by averaging several isolated sources, such that the extracted PSF
is independent of fitted analytical functions. This is particularly crucial,
as the Keck AO PSF cannot be characterized by azimuthally symmetric functions.
The emploid version of the code is limited to non-spatially varying PSFs.
The radial extent of the PSF was set
for each field and filter by the radius where the PSF structure was 
dominated by background noise, and is listed in Tab.~\ref{obstab}.
In the cluster core, $\approx 10$ sufficiently bright and isolated sources
could be found to define the average PSF. In the east fields 
(see Fig.~\ref{mosaic}), the rapidly
decreasing density of bright stars limited the PSF 
template selection to 5-7 sources on field east1, and 3-5 sources on east2.
The final PSF for each field was created iteratively after selection and 
subtraction of neighboring sources from each image. Three iterations were performed
before the final PSF was extracted. The average empirical PSF then served 
as the template to fit simultaneous photometry for all sources in each frame.

In addition to the deep, combined images, the reduced frames were divided into
three qualitatively comparable subsets and 
combined into three auxiliary images to estimate the photometric and 
astrometric uncertainties. 
The auxiliary frames are composed from the list of all selected
images sorted by Strehl ratio. From this quality-sorted list, every 
third frame enters auxiliary frames 1, 2, and 3, respectively, to ensure 
the same quality in all three auxiliary frames. The random dither pattern
with three images per pointing position and small dither offsets
additionally ensures that
image distortions are sampled in the same way in all three auxiliary frames.
Starfinder positions and photometry were derived for 
point sources on the auxiliary images in the same way as for the deep data sets. 
As a first pass at rejecting faint, spurious detections caused by the varying
halo structure of the PSF or fluctuations in the background, sources in 
$H$ and $K'$ were required 
to be detected in at least one auxiliary frame in addition to the deep image 
in order to be included in the final photometry table.
In addition, the auxiliary frames were used to estimate the photometric 
and astrometric uncertainties.
In $H$ and $K'$, the uncertainties are dominated by crowding effects. 
The photometry of stars near bright sources
is sensitive to the Strehl ratio of each image, and hence their individual
measurements on the three auxiliary images vary more than for isolated sources. 
This crowding effect is not evident in the photometric uncertainties delivered 
during PSF fitting by starfinder, but is 
well reproduced in the rms uncertainties from repeated measurements on the 
auxiliary images.
The astrometric and photometric uncertainties, shown in Figs.~\ref{photuncer}
and \ref{posuncer},
are derived from the standard deviation (rms) of the positions and fluxes 
on the auxiliary images and the main image, 
and hence from the repeatability of each measurement. 
In $L'$, residual thermal background emission caused the sensitivity 
of the auxiliary images to be significantly lower than the sensitivity 
of the deep image, such that $L'$ uncertainties are most affected
by spatial sensitivity variations and not by crowding. 
The limited photometric sensitivity 
of the auxiliary frames restricted the number of $L'$ sources 
matched with the photometry of the $H$ and $K'$ images. 
With the goal of including the fainter $L'$ sources detected only in the deep 
combined image in the final catalogue, all $L'$ starfinder detections were 
matched with the $HK'$ photometric table. When sources were detected in the 
auxiliary frames, the $L'$ photometric uncertainties
were derived from the standard deviation of the repeated measurements.
For faint sources not detected in the auxiliary frames, the starfinder
signal-to-noise ratio (S/N) was used to estimate the photometric uncertainty.
As PSF fitting does not account for background fluctuations, the starfinder
S/N still appears to underestimate the photometric uncertainty in some cases.
A comparison of the rms photometric uncertainties derived from the auxiliary frames 
and the starfinder S/N indicates that the starfinder S/N may underestimate 
the true $L'$ photometric uncertainty by up to a factor of 4. These smaller 
uncertainties have a negligible effect on the final photometric uncertainty 
of each source, which is dominated by the uncertainty in the zeropoint 
(see Table~\ref{zpttab}). The final $HK'L'$ table thus contains
as many sources as possible, while avoiding $L'$ artefacts with the requirement
that any $L'$ source be detected in all three filters. 
This procedure resulted in a clean list of 380 $HK'L'$ detections and 
1033 additional $HK'$ detections for the two-color analysis (Table \ref{fulltable}).

\subsubsection{$HK'$ photometric calibration}

The $HK'$ photometry was calibrated with reference to the VLT/NAOS-CONICA (NACO)
observations obtained in 2002 (Stolte et al.~2002, 2005, see also Stolte et al.~2008),
covering a $26^{''}\times 27^{''}$ FOV around the cluster center with a spatial resolution 
similar to the Keck/NIRC2 observations (Tab.~\ref{obstab}). The absolute calibration 
zeropoint for the NACO photometry is provided by lower resolution ($0.2^{''}$)
HST/NICMOS calibrations presented in Figer et al.~(1999). 
Five available 2MASS sources were used to 
check that the calibration with respect to the NICMOS filter system did 
not display systematic offsets with respect to the standard 2MASS $HK_s$ system. 
As in our previous work, no color terms were found (Fig.~\ref{colorterms}). 
In order to verify that 
the use of a spatially invariant PSF did not alter the photometric performance
across the field, photometric residuals were tested independently in the x, y and radial
directions. A possible weak zeropoint variation of less than $\pm 0.1$ mag, 
comparable to the standard deviation of $\sigma_{\Delta K} = 0.10$ mag in $K'$ and 
$\sigma_{\Delta H} = 0.15$ mag in $H$ in a sample of 76 calibration sources, 
may be present over the maximum radial extent of $5.5^{''}$ in the NIRC2 
$H$ and $K'$ data sets when compared to both NICMOS and NACO, 
but the linear fit indicates less than one sigma significance. 
As the variance is within our formal zeropoint uncertainties (Table~\ref{zpttab}),
we apply a constant zeropoint across the field. With increasing distance 
from the field center, anisoplanatic effects are expected to affect the 
natural guide star observations with NACO most severely, while more modest effects 
are expected for the Keck LGS observations since the artificial guide star was 
pointed at the center of each $10^{''}\times 10^{''}$ field of view.
This may explain the slight differences of 0.03-0.10 between the zeropoints 
of the NIRC2-NIRC2 and the NIRC2-NACO calibrations.
In order to achieve a uniform photometric calibration across the NIRC2 mosaic 
area, we therefore apply a two-step correction process.
First, absolute photometric calibration is obtained for the central core region
from the comparison with NACO, and relative calibration of fields at larger
distances from the guide star is then derived with respect to the cluster core.
The core field is close to the guide star used both as the natural GS with NACO 
and as the tip-tilt source with NIRC2, such that anisoplanatic effects at 
increasing guide star distances are minimized (see Fig.~\ref{mosaic}). 
For the core field, we use only the innermost $5^{''}$
region within $250 < x,y < 750$ NIRC2 pixels to calibrate against NACO $HK_s$.
The high stellar density in this area provides 56 sources with 
$\sigma_{Ks} < 0.1$ and $\sigma_H < 0.15$ mag for the absolute calibration,
and zeropoint variations over this small area were confirmed to be negligible.
The zeropoint offsets of the adjacent fields,  
``east1'' and ``lead'' were derived with respect to the calibrated core field, 
and ``east2'' and the intermediate proper motion field ``east1.5'' 
were calibrated against ``east1'' (see Table \ref{zpttab}). Only sources 
with relative photometric uncertainties $\sigma_{HK'} < 0.1$ mag, 
as derived from the auxiliary images (see above), were used as calibration sources.
The halo field does not overlap with the core field, but its proximity 
to the guide star ensures that anisoplanatic effects are avoided.
Hence, the halo field $HK'$ photometry was calibrated against the NACO field.
The average zeropoints after airmass correction were 25.219 in $H$, 24.578 in $K'$.

\subsubsection{$L'$ photometric calibration}
\label{lzero}

%
%
%
%

In $L'$, where no previous data exist, absolute calibration was achieved using 
$L'$ observations of the central parsec in the Galactic center (GC), 
where numerous previous
long-wavelength observations are available (see Appendix, Tab.~\ref{gctab},
and for reference, see also Blum et al.~1996, Cl\'{e}net et al.~2001, 2004).
The central parsec was observed during the same night and at the same airmass 
as the Arches core field. 
The surrounding Arches fields were subsequently calibrated relative to the 
core field. The same observational 
setup was employed for both the Arches and the GC observations. Taking into 
account airmass variations, we first derived the zeropoint for the GC data by 
referencing against calibrated NGSAO observations reported in Appendix A.
The final calibration was confirmed by matching with both the Blum and 
Cl\'{e}net tables. 
This procedure ensures that the zeropoint offset between the $L'$
($\lambda_c = 3.75 {\rm \mu m}$), used by Cl\'{e}net et al.~(2001) and for 
our data, and the $L$ ($\lambda_c = 3.45 {\rm \mu m}$) filter used in 
Blum et al.~(1996) was accounted for. The difference between $L$ and $L'$ is 
a constant offset in this case, as sources in the GC sample
as well as in the Arches cluster have comparable line of sight extinctions 
(see Appendix B for details on the filter differences and the extinction law)
\footnote{Note that Viehmann et al.~(2005) provided the most recent
photometry in the central parsec, however, these data unfortunately display an
extensive scatter in $K'$ as well as $L'$, such that they could not be used as
calibration sources.}. The uncertainty in the zeropoint was estimated from 
variations of the zeropoints of individual images throughout the observation 
sequence.
The derived GC $L'$ zeropoint of $23.30 \pm 0.10$ mag was used to calculate 
magnitudes for the Arches core field after airmass correction. 
Possible temperature or atmospheric variations leading to a varying 
thermal or sky background during the two hours of $L'$ 
observations cause the remaining uncertainty of 0.10 mag in the zeropoint.
When the calibrated photometry was compared to the location of the reddening vector 
in the $H-K', K'-L'$ two-color diagram, the majority of sources was located below
the reddening line. This is unexpected, as the cluster population is dominated by 
main sequence stars, and the field population is dominated by field giants.
The intrinsic colors of main sequence stars are close to zero in the standard
magnitude system, such that the reddening vector also passes through the origin
in the two-color diagram, and main sequence stars behind substantial foreground 
extinction, such as the Arches members, are expected to scatter around this vector
according to their photometric uncertainties.
A shift of $\Delta L' = +0.1$ mag allowed for both cluster stars and 
background sources in the Arches field to follow the extinction vector 
(see Sec.~\ref{lsec} below), as expected. 
Such a zeropoint shift can be 
explained by the difference in the average extinction between 
the GC calibration sources, $<A_V> = 33.1$ mag  and Arches cluster members,
$<A_V> = 25.3$ mag. $\Delta A_V = 7.8$ mag corresponds to a difference 
of $\Delta L' = 0.12$ mag when calibrating $L'$ with respect to $L$ data
(see appendix B for details). 
This offset is, however, also consistent with the zeropoint variation 
across the two hour $L'$ GC observing sequence due to 
thermal background variations across the night.
As the offset is within the zeropoint uncertainties, it was applied to 
the absolute $L'$ calibration.

All adjacent images were calibrated with respect to the core,
and as in the case of $HK'$, the zeropoint uncertainties were calculated based on 
the standard deviations of the photometric residuals of the calibration sources 
after the zeropoint was applied.
In the halo field, four $L'$ acquisition exposures with total exposure times of 30s 
each were obtained with one third of the acquisition images overlapping the core field,
and two thirds overlapping the lower-density halo field. The acquisition exposures were
combined to provide a calibration field for the halo field $L'$ photometry.
The zeropoint of the acquisition image was obtained with respect to the core,
and the halo field was calibrated with respect to the acquisition image, consecutively.
The reported systematic zeropoint uncertainty in $L'$ for this field is comprised
of the standard deviations of the photometric residuals of the acquisition and 
halo fields.

\subsubsection{Zeropoint uncertainties}

Zeropoint uncertainties are estimated from the standard deviation in the photometric
residuals of calibration sources after the zeropoint was applied (Table \ref{zpttab}), 
and are included as systematic uncertainties in all reported photometric uncertainties. 
Where several fields had to be used consecutively to obtain the absolute zeropoint,
the standard deviations from each field were added in quadrature to obtain the 
final zeropoint uncertainty.
The zeropoint uncertainties in east1 comprise the quadrature sum of the core 
NIRC2-NACO residuals (first row in Table \ref{zpttab}) 
and east1 residuals ($\sigma_H = 0.065$, $\sigma_{K'}=0.063$, $\sigma_{L'}=0.054$).
The intermediate field east1.5 was only observed in $K'$ for proper motion membership. 
The zeropoint uncertainty contains the standard deviation from all consecutive 
calibrations (i.e., the zeropoint uncertainties of $K'$ core and $K'$ east1), 
added in quadrature to the measured standard deviation 
in $K'$ east1.5 ($\sigma_{K'}=0.086$) after calibration.
The east2 $HK'L'$ zeropoints were derived from field east1.5 in $K'$, and field east1 
in $H$ and $L'$, and the uncertainties contain the uncertainty in the core, east1 and east1.5 
calibration in addition to the residual standard deviation in east2 after calibration 
($\sigma_H = 0.088$, $\sigma_{K'}=0.073$, $\sigma_{L'}=0.030$).
The leading field $HK'L'$ zeropoints were derived from overlap with the core field, 
and the uncertainties include the quadrature sum of the core zeropoint uncertainties and 
the standard deviation in the residuals of the leading field after calibration 
($\sigma_H=0.086$, $\sigma_{K'}=0.079$, $\sigma_{L'}=0.100$).
The halo field $HK'$ photometry was zeropointed against the NACO $HK$ catalogue. 
The zeropoint uncertainty is the residual standard deviation in the calibration sources 
after the zeropoint was applied. 
$L'$ photometry on the halo field was calibrated in two steps with respect to 
an aquisition exposure with significant overlap to the core and halo fields.
The $L'$ zeropoint uncertainty contains the uncertainties in the core and aquisition image 
calibrations in addition to the residual standard deviation in the halo field after the 
zeropoint was applied
($\sigma_{L'core}=0.100$, $\sigma_{L'aqu}=0.064$, $\sigma_{L'halo}=0.054$).
Final zeropoint uncertainties from this procedure are tabulated in Table \ref{zpttab}.

\subsubsection{Photometric uncertainties and completeness limits}

Fig.~\ref{photuncer} shows the relative photometric uncertainties in each field, 
color-coded by filter. As discussed above, the relative uncertainties were derived
from the repeatability of each photometric measurement from the three auxiliary 
images in each field and filter. These uncertainties illustrate the filter dependence
in the photometric accuracy and the uniformity between the different observations.
The absolute uncertainty is dominated by the systematic zeropoint uncertainty,
which appears as a constant offset in these plots and is not included here, such
that the relative measurement uncertainties can be directly compared.
The dashed lines indicate the peak in the observed 
luminosity functions (LF) as a proxy for the completeness limit. 
In the crowding limited field of the Arches 
core, the non-detection of sources in high-density regions 
shifts the peak of the LF to brighter luminosities. While numerous stars can 
thus be detected at magnitudes fainter than the LF peak, the sample 
beyond this limiting magnitude will be incomplete.
In field east1 ({\sl middle panel}), where only 10 frames with good AO 
correction were obtained, the $K'$-band photometry is significantly
shallower. Where sources were detected in the overlap regions between
east1 and the cluster core or east2, sources were preferentially selected
from the more complete core and east2 photometric samples.
The last column in Tab.~\ref{obstab} contains the peak of the luminosity
function (LF) in each field and filter, which we identify as the completeness
limit. In the crowding-limited cluster core, the $HK'L'$ completeness limits are 
19.0, 19.5, 14.0, and in the remaining fields the average limits are 
20.5, 19.5, 14.5 in $HK'L'$, respectively.
The total numbers of unique sources 
detected in all three filters in the core, east1, east2, leading, and halo west 
fields are 173, 32, 39, 73, and 63, respectively. These $HK'L'$ samples are limited
by $L'$ completeness, and faint sources without $L'$-band excess will not be included
in this sample. For the derivation of the excess fraction, it is crucial
to have a complete main sequence reference sample (see Sec.~\ref{excsec}).
Hence, we also define a sample of $HK'$ detections down to the $H$-band 
completeness limit of 19.0 mag in the most crowding-limited cluster core.
The $HK'$ sample includes 261, 63, 75, 108, 61 unique detections with 
$H < 19.0$ mag
in the core, east1, east2, leading, and halo west fields, respectively.
Note that these samples are not yet selected for cluster membership, 
but include the full number of detections in our fields.

\subsection{Proper motion membership}
\label{pmsec}

Following the procedures detailed in Stolte et al.~(2008), the $K'$-band
catalogue was matched with VLT/NAOS-CONICA observations, providing a 
time baseline of 4.3 years. 
As different fields were observed over several years with NIRC2
during different atmospheric conditions,
the geometric transformation of each NIRC2 $K$-band image was derived 
with respect to the larger NACO image individually. The transformations
are derived iteratively, with the first pass determined from bright, 
likely cluster members, and the second pass from a preliminary selection
of member candidates of all magnitudes to improve spatial coverage and 
thus the quality of the transformation. 
This procedure ensures that variations due to adaptive
optics performance and airmass differences are removed. Note that this 
also implies that internal cluster motions, which are expected to be below
our NIRC2-NACO proper motion accuracies, are also minimized and cannot 
be detected with this method. A comparison of NIRC2-NIRC2 epochs leading
to the determination of internal motions will be the topic of a forthcoming 
paper (Clarkson et al.~2010, in prep.).
In addition to the core members determined in Stolte et al.~(2008), 
we have included sources with membership information on field east1,
as well as on the leading and halo-west fields,
where overlap with the NACO data set is available (see Fig.~\ref{mosaic}).
A total area of 442 square arcseconds was observed in $HK'L'$, of which 
312 square arcseconds are also covered with NACO. 
Fields east1 and east2 did not overlap entirely with our NACO observations.
The intermediate field east1.5 covering the missing portion of the 
eastern part of the cluster was re-observed with Keck/NIRC2 in July 2008 
to obtain second-epoch $K'$ images (see Tab.~\ref{obstab}), which increases 
the area with proper motion measurements to 392 arcseconds squared. Hence,
90\% of the $HK'L'$ coverage has proper motion information. 
The standard deviation in the proper motions of cluster members derived for
fields east1/east2 using east1.5 as proper motion reference over a two-year 
baseline with NIRC2 was comparable to the astrometric residuals
in the four-year NACO-NIRC2 comparison, such that no distinction between the 
two data sets had to be made. 

Only stars 
brighter than the core $H$-band incompleteness limit (peak of the LF), 
$H < 19$ mag, were included in the fit, as these stars 
represent the main sequence sample in the analysis below.
The membership selection of stars brighter 
than $H=19$ mag is illustrated in Fig.~\ref{xymove}. The gaussian fit
to the strong concentration of stellar motions in this sample yields a 
proper motion dispersion of $\sigma_{gauss} = 0.65$ mas/yr, and stars within
2-sigma of the peak, or a proper motion of less than 1.30 mas/yr
relative to the average motion of the Arches cluster, 
were considered cluster member candidates (hereafter 'proper motion members'). 
In the following
sections, members and non-members will be distinguished on the 
basis of this selection criterion. As a consequence of the 
different resolutions in the two data sets, the proper motion 
table is not complete in the cluster core. Faint sources or 
close neighbours resolved with the 53 mas resolution 
obtained in the NIRC2 $K'$-band images were not always resolved 
with the 84 mas resolution obtained with NACO. 
This leads to 16 sources without proper motion information in the 
cluster core, one of which has $L'$-band excess (see Sec.~\ref{excsec}).
The membership status of the 15 sources without $L'$ excess is uncertain,
and therefore these are not considered members. The detection 
of $L'$-band excess suggests that this one sources belongs to the cluster,
such that this source is included, albeit seperately, in the discussion below.

This procedure provides membership information for 479 (84\%) sources
of the 568 $HK'$ detections with $H < 19$ mag, including 335 cluster 
members. The final $HK'L'$ catalogue, which is limited by the $L'$ detection
threshold of $\sim 14$ mag, contains 362 detections with
$H < 19$ mag, of which 331 (91\%) have proper motion information and 
235 are cluster member candidates within the 2-sigma member 
selection criterion.

\subsection{VLT/SINFONI K-band spectroscopy}
\label{specsec}

VLT/SINFONI integral-field spectroscopy was obtained 
in the core of the Arches cluster between 2006 May and 2006 July.
The medium pixel scale with $50 \times 100$ mas/pixel
spaxels delivered a field of view of $3^{''} \times 3^{''}$, and
the $H+K$ grism yielded a mean spectral resolution of 1750.
Two fields 
in the cluster core were observed, and the total integration time was 2 hours 
per field split into 4 spectral cubes of 30 min each. Each cube consisted of
six individual 300 s exposures obtained with an ABBA dither pattern
alternating between science and sky frames pointed to a dark region 
near the cluster to account for variable sky levels in the near-infrared. 
The spectra were extracted from the pipeline-reduced data delivered
by ESO, which includes standard sky subtraction, flat fielding, masking of 
hot pixels, and wavelength distortion correction and calibration.
The wavelength scale was rebinned to an equal wavelength spacing of
0.0005$\micron$/pixel, providing a nominal
spectral resolution of 2000 at 2.2$\micron$ or 140 km/s per wavelength element.
From telluric lines, a spectral resolution of 1900 is measured near 
2.2$\micron$, or 160 km/s (FWHM). 
Telluric correction was achieved either with a G2V or a B3V calibrator
observed during the same night. In the case of the G star, the rebinned, 
high-resolution solar spectrum available from KPNO was used to remove 
intrinsic stellar features, while for the B3V calibrator, the broad
hydrogen lines were removed using template spectra from the $K$-band atlas
of Hanson et al.~(1996). In order to correct for flexures in the SINFONI
instrument, the telluric spectra were cross-correlated with bright 
stars in the science cubes, sub-pixel shifted and rebinned to the 
original wavelength scale. This procedure resulted in a clean removal 
of almost all telluric lines, except for one residual line at 2.355$\micron$,
visible as a strong contaminant at the location of the third CO overtone 
in all the spectra. This line might originate locally near the telescope
or inside the instrument, as telluric sky correction did not allow for 
its removal in either of our science data cubes. Three sources with
$L'$ excess (see Sec.~\ref{lsec}) were located in two SINFONI fields.
For these three $L'$-band excess sources with $K=15.9, 15.2,$ and 15.5, 
the spectra of three to four data cubes were added to provide the final, 
extracted spectrum of each source with signal-to-noise ratios (S/N) 
of 40, 65, and 60, respectively, in the $K$ continuum near the 
CO emission band heads.

\section{Results}
\label{excsec}

\subsection{Photometric and spectroscopic evidence for circumstellar disks 
in the Arches cluster}
\label{diskexcsec}

\subsubsection{Identification of $L'$-band excess sources}
\label{lsec}

Near-infrared excess at wavelengths longward
of 3 microns can be caused by circumstellar dust emission.
At different stages of stellar evolution, the emitting molecular 
material can be found in different geometries around the central star.
In very young main sequence stars, dust emission 
is attributed to the existence of a circumstellar disk,
assumed to be the remnant disk from the accretion phase
of the star. A remnant disk is less likely for high-mass stars
earlier than spectral type B3V, 
where native material is evaporated by UV radiation from 
the central star on timescales of one to a few Myr (Richling \& Yorke 1997, 
Hollenbach et al.~2000). 
At later stages of high-mass stellar evolution, in particular
during the Wolf-Rayet (WR) phase, the winds of 
high-mass stars can produce massive, dusty envelopes, in which 
the stellar light is also reprocessed and emitted at near- to 
mid-infrared wavelengths (see Crowther 2007 for a review). 

The $H-K'$, $K'-L'$ two-color diagram (Fig.~\ref{ccd_hkl}) 
allows us to distinguish sources with enhanced foreground extinction 
from infrared excess sources. The solid line indicates
the reddening path of an A0 star using the extinction law as measured towards 
the GC (Rieke \& Lebofsky 1985). In Fig.~\ref{ccd_hkl} (left), 
cluster members are plotted in light blue, while non-members are displayed in 
red. Black dots indicate stars without membership information.
The bulk of the Arches members follow the reddening vector, as expected from the 
range of extinctions of $22 < A_V < 30$ mag observed towards 
the cluster center (see also Stolte et al.~2002, Espinoza et al.~2009).
Non-members are preferentially found at higher extinctions, but still 
concentrated around the reddening path. 
The proper motion member sample of Arches cluster stars is shown 
in the right panel of Fig.~\ref{ccd_hkl}. The efficiency of the 
membership selection is evidenced by the lack of sources above a 
foreground extinction of $A_V \sim 33$ mag along the reddening vector.
A significant sample of extremely red cluster members with $K'-L' > 2.3$ mag
is found to the right of the reddening vector, clearly distinct from both the 
cluster main sequence and the field populations. For these excess sources, 
we have to investigate whether they are evolved high-mass stars, 
or whether the excess emission stems from a circumstellar disk.

Spectroscopically classified evolved, high-mass stars (Martins et al.~2008) 
are indicated as crosses and asterisks in the two-color diagram. 
While O4-6 supergiants cluster
close to the main sequence population, nitrogen-enriched WN8-9h
stars occupy the locus of classical T Tauri stars (CTTS, Meyer et al.~1997), 
which represents characteristic colors for young, disk-bearing stars. 
The $L'$-band excess of these Wolf-Rayet stars is, however, not caused by 
disk emission, but originates in extended dusty envelopes produced by their 
strong stellar winds. 
Three of the brightest cluster members ($K < 11.5$ mag) not covered 
by the Martins et al.~(2008) spectroscopic survey of evolved stars 
display colors comparable to the identified WN8-9h stars, indicating 
that these sources have also started their post-main sequence evolution.
These contaminants in the two-color diagram are found only in massive star clusters, 
and are not a source of uncertainty in the disk fractions measured
in lower-mass star-forming regions. 
For the purpose of deriving the disk fraction in the Arches cluster 
(Sec.~\ref{diskspecsec}), evolved stars with $K' < 14$ mag are 
excluded from the excess sample representing circumstellar disks. 

Several very red cluster members with excesses of at least $0.3 \pm 0.02$ mag redder 
in $K'-L'$ than even the {\sl evolved} cluster population stand out in the two-color 
diagram at $K'-L' > 2.3$ mag.
While the spectroscopically studied sources in Martins et al.~(2008) 
comprise the brightest part of the cluster main sequence with likely progenitor
masses above $60\,M_\odot$, the comparison with the CMD (Fig.~\ref{cmd})
reveals that the extreme excess sources are faint. With $K'$-band magnitudes 
of 14.5 to 18 mag (Tab.\ref{hkltab}), their stellar masses 
are expected to be substantially lower than Wolf-Rayet stars (Sec.~\ref{masssec}).
As the spectra of three of these sources with large $L'$-band 
excess display strong CO bandhead emission lines (Sec.~\ref{diskspecsec}), 
their $L'$-band excesses most likely originate in circumstellar disks.

Formally, we define disk sources as cluster members with $K'-L'$ colors 
too red to be reddened main sequence stars, and that are too faint to 
be Wolf-Rayet stars ($K > 14$ mag).
To distinguish sources with significant $K'-L'$ excess from the cluster
main sequence, 
we determine the standard deviation in the main sequence population from  
the rms in the $K'-L'$ and $H-K'$ color of stars with 
$K'-L' < 1.6$ mag to be $rms_{KL} = 0.11$ mag and $rms_{HK} = 0.14$ mag.
The 2-sigma ellipse with a major and minor axis of twice the derived 
standard deviations in $H-K'$ and $K'-L'$, respectively, is shown in the 
two-color diagram in Fig.~\ref{ccd_hkl}.
Note that the apparent circular shape of the ellipse is a consequence of 
the unequal axis spacing. The tangent lines parallel to the reddening vector 
(dashed lines) indicate the color regime where cluster stars with higher 
foreground extinctions can be found. The color variations of sources 
above the reddening path are dominated by their photometric uncertainties.
These randomly scattered sources above the reddening path are enveloped 
by the dashed lines, indicating that all main sequence members without excess 
both above and below the reddening path are confined between these 2-sigma limits. 
Hence, we define stars significantly beyond these 2-sigma limits as excess sources.
Specifically, sources to the right of the lower dashed line in Fig.~\ref{ccd_hkl}
are required to meet the following criteria: \\

\noindent
$(K'-L') - \sigma_{K'-L'}\,>\,(H-K')/m_{ext}\,+\,0.37$ \\
\\
$(H-K') + \sigma_{H-K'}\,< $ 

\hspace*{2cm} $m_{ext}\,*\,(K'-L')\,-\,m_{ext}\,*\,0.37$ \\

where the slope of the Rieke \& Lebofsky (1985) extinction law is given by 

$m_{ext} = (A_H-A_K)/(A_K-A_L) = (0.175-0.112)/(0.112-0.058) = 1.1667$. \\

The y intercept is determined by the shift of the reddening vector to the 
tangent of the 2-sigma ellipse, $\delta_{2\sigma} = 0.37$. 
$\sigma_{H-K'}$ and $\sigma_{K'-L'}$ denote the photometric color uncertainties 
of each source.
These requirements imply that sources are counted as having an excess when 
their colors are significantly to the right of the 2-sigma reddening band 
in $K'-L'$ and significantly downward of the 2-sigma reddening band in $H-K'$.
Fig.~\ref{ccd_hkl} (right) 
shows the selection of significant excess members as diamonds. 
The bulk of the excess sample is shown in the insert in Fig.~\ref{ccd_hkl} 
(right) with photometric uncertainties, illustrating that the uncertainty 
in the colors of each excess source cannot account for their observed 
distances from the main sequence locus and the reddening band. 
With these selection criteria, 24 $L'$-band excess sources are 
detected with $K' > 14$ mag. In addition, six stars with $K' > 14$ mag 
are located close to the CTTS locus, but are consistent with a locus inside
the reddening band when their photometric uncertainties are taken into account.
These sources can either be main sequence cluster members suffering enhanced 
foreground extinction along the line of sight or sources with weaker $K'-L'$ 
excesses. As the nature of these stars cannot unambiguously be determined 
at this point, they are not included in the excess sample.
The photometric properties of the $L'$-band excess sources are 
summarized in Table~\ref{hkltab}.
Of the 24 sources with infrared excess, 23 have measured proper 
motions (red diamonds in Fig.~\ref{xymove}), while one star
is not resolved in the NACO reference image due to its proximity to 
a bright neighbor. One source, located at the edge of the NIRC2 field 
of view, is formally a non-member, but its positional uncertainty 
in the x direction is a factor of five higher than the mean uncertainty 
in the excess sample and 
does not allow for a final conclusion on its membership. The second formal
non-member is blended with a source of comparable brightness, and the 
two sources are not well resolved in the NACO $K$-band reference image.
All 21 infrared excess sources with reliable proper motion measurements
are proper motion members of the Arches cluster. 
 It is extremely unlikely to find 21 sources with specific
 colors to be cluster members simultaneously.
 The region in the proper motion plane covered by the majority of the 
 field stars amounts to $5\times 10 {\rm mas/yr}^2 = 50 {\rm (mas/yr)}^2$, 
 while the excess sources cover an area of $2 \times 2 {\rm (mas/yr)}^2$, 
 and the full 2-sigma circle is exactly 
 $\pi \cdot 1.28^2 = 5.15 {\rm (mas/yr)}^2$. Hence, the likelyhood to 
 find a single field source inside the 2-sigma candidate circle is
 5.15/50 = 0.10, and the likelyhood to find 21 sources in this area 
 of the proper motion plane simultaneously is only $(0.1)^{21} = 10^{-21}$.
Given the uncertain proper motion of the 3 remaining sources with 
$L'$-band excess colors, we consider all 24 sources members
of the cluster.

Within the completeness limit in the cluster core, $H < 19$ mag,
the $L'$-band excess cluster members 
occupy a magnitude range of $16.6 < H < 19.0$, 
and span a color range of $2.03 < K'-L' < 3.38$ (Tab.~\ref{hkltab}).  
The cluster main sequence population covering the same $H$-band magnitudes
displays a mean color of $K'-L' = 1.42 \pm 0.11$ mag (rms).
With a mean color of $K'-L' = 2.73 \pm 0.36$ mag (rms), the 19 $L'$-band 
excess members with $H < 19$ mag are separated from the cluster main sequence by 
$\Delta (K'-L') = 1.31^{+0.65}_{-0.70}$ mag, and hence are redder than 
the mean main sequence color by at least 0.6 mag
(see Fig.~\ref{cmd}, right panel). The comparison with 
the $H-K', K'$ CMD shows that the excess sources are not
revealed by $HK$ photometry alone. In particular, the bluest sources
with $L'$-band excess emission overlap in the $H-K'$, $K'$ CMD with the 
color-magnitude location of the red clump in the nuclear bulge 
($K' > 15.3$ mag, $H-K' > 1.8$ mag, 
for $A_K \ge 2.4$ mag, $M_K = -1.6$ mag, Alves 2000). 
The empirical location of red clump stars stands out in our non-member 
sample as a clustering at $H = 17.5$, $H-K = 1.87$ mag and
is marked with a red box in the CMDs. The ambiguity between excess sources 
and red clump stars is resolved in $K'-L'$, emphasizing the importance of $L$-band
observations to derive disk fractions in young star clusters. 
The completeness limits indicate that the detection of fainter excess candidates 
is limited by the current $L'$-band completeness limit of $L' = 13.5-14.5$ mag
in each field.

In summary, we detect 24 sources with significant $L'$-band excess, of
which all 21 excess sources with reliable membership information are 
proper motion members of the Arches.

\subsubsection{Spectral types and stellar masses of disk host stars}
\label{masssec}

In the Arches cluster, photospheric flux measurements of excess sources 
at visible wavelengths are prohibited by the large foreground extinction 
towards the GC. Hillenbrand et al.~(1992) provide evidence that
photospheric emission dominates the SED of Herbig Be stars below 
$1.2 \mu$m, i.e.~$J$-band. Even in $J$-band observations, however, 
most fainter sources are veiled by the foreground extinction of 
$22-30$ mag towards the Arches (Stolte et al.~2002). 
The $H-K', H$ CMD (Fig.~\ref{cmd_disks}) shows that the near-infrared
excess in $H-K'$ is substantially smaller than in $K'-L'$.
The Arches excess sources display on average about 0.5 mag less $H-K'$ 
excess than the Hillenbrand Herbig Be sample, shown as box points. 
A lower excess at smaller wavelength indicates that the hot, inner disk
rim becomes increasingly depleted at these ages, possibly due to the 
existence of a growing inner hole. As a consequence, the $H$-band flux
of the excess sources will be closer to the expected main sequence 
flux than in younger excess populations.
Hence, we use the $H$-band magnitude to derive an approximate mass
range for the excess population. 

The $H$-band magnitudes are compared to a solar-metallicity, 2.5 Myr 
main sequence isochrone from the standard set of Geneva stellar evolution 
models (Lejeune \& Schaerer 2001).
The foreground extinction to each excess source (Tab.~\ref{hkltab}) is measured
from the  $H-K', K'-L'$ color of the four nearest cluster members without excess
emission using a Rieke \& Lebofsky (1985) extinction law. 
The variation in visual extinction ranges from 22.7 to 27.9 mag, 
or $A_K = 2.5$ to 3.1 mag. The variation between individual neighbouring sources 
used to estimate the average extinction on the line of sight to each excess source 
is large, yielding standard deviations of up to $\sigma_{AV} = 2.4$ mag.
The bright stars concentrated towards the cluster center have blown a cavity
evidenced by a lower reddening in the cluster core (Stolte et al.~2002).
The mean foreground extinction towards the cluster center is therefore determined 
from the upper main sequence,
$12 < K' < 15$ mag and $1.2 < K'-L' < 1.5$ mag, which is least biased by 
variable extinction (see Fig.~\ref{cmd}), to $A_{V\,ms} = 25.2 \pm 1.8$.
This value is identical to the mean foreground extinction of $A_V = 25.3 \pm 1.4$ mag 
of the excess sample. As substantial small-scale spatial variation
leads to the large uncertainties in the extinction of each excess source,
we use the average extinction towards the excess sample for stellar mass computation. 
The extinction of $A_V = 25.3$ mag and a distance modulus
of $DM = 14.52$ mag (8 kpc) are applied to the isochrone $H$ magnitude.
No color transformation to the Mauna Kea system was applied, 
as the difference is smaller than 0.05 mag
(Carpenter 2001, Hawarden et al.~2001).
The 19 excess sources above the core $H$-band completeness limit
are observed within a magnitude range of 
$16.7 < H < 19$ mag, corresponding to a mass interval of
$15 > M > 5\,M_\odot$. On the zero-age main sequence (ZAMS), 
these masses correspond to spectral types B1V - B6V.
Two $L'$ excess sources with
$H = 19.8$ and $H = 20.7$ are not included in the complete disk 
sample, and the comparison of their faint $H$-band magnitudes with 
a Geneva 2.5 Myr isochrone indicates that 
disk-bearing A stars (Herbig Ae stars) with masses down to at least
$2.2\,M_\odot$ exist in the Arches cluster.
Likewise, when the maximum foreground extinction uncertainty 
of $\sigma_{AV} = 2.4$ mag, corresponding to $\sigma_{AH} = 0.42$ mag, 
is taken into account, the full allowable mass range could be as large as 
$1.8 < M/M_\odot < 18$.
A similar lower mass limit is obtained by allowing for 
an unaccounted residual $H$-band excess, which might influence
the luminosities of the reddest excess sources in the sample, 
which have $H-K'$ colors similar to the Hillenbrand Group I star/disk
systems. 
Characteristic $H$-band excesses in very young Herbig Be stars are 
in the range 0.3-1 mag (see, e.g., Fig.~2 in the Herbig Ae/Be survey
of Hern\'{a}ndez et al.~2005). If we allow for a maximum $H$-band excess 
of 1 mag in the two reddest excess sources, 
these sources could also have masses as low as $3\,M_\odot$.
At the same time,
a small fraction of the disks should be seen edge-on, and for these
sources local extinction in the optically thick disk causes additional 
dimming of the $H$-band flux. 
A circumstellar visual extinction of 10 mag corresponds
to 1.7 mag in $H$ (Rieke \& Lebofsky 1985), 
such that some of the excess sources could be 
intrinsically brighter with masses in excess of $20\,M\,_\odot$.

In summary, the $H$-band luminosities of the excess population 
indicate that several A5V to B0V stars within a maximum mass 
range of $2 < M < 20\,M_\odot$ feature disks in the Arches cluster.

\subsection{Spectroscopic evidence for circumstellar disks}
\label{diskspecsec}

\subsubsection{CO bandhead emission and evidence for disk rotation}
 
$K'$-band spectra of 3 of the 24 $L'$-band excess sources
are shown in Fig.~\ref{spectra}. The red lines are the extracted 
spectra of the excess sources from the SINFONI data cubes, while 
the black spectra are bright stars located in the immediate vicinity
of each excess source. All 3 excess sources display strong CO bandhead
emission with the first and second overtones at 2.29$\mu$m and 
2.32$\mu$m, respectively. The first two overtones are detected 
at high S/N ratios of 34-118, where the first overtone is always 
stronger by approximately a factor of two as compared to the second
overtone emission. 


CO bandhead emission 
in high-mass YSOs with spectral types O6-B5 are observed in 
NGC 3576 at ages below 1-2  Myr 
(Figuer\^{e}do et al.~2002, Blum et al.~2004) and in the 
very young $< 1$ Myr cluster NGC\,6618 in M17 (Hanson et al.~1997, 
Hoffmeister et al.~2006). 
There is increasing evidence that the CO bandhead emission
originates in rotating disks.
In agreement with previous disk models (Chandler et al.~1995, 
Bik \& Thi 2004, and Blum et al.~2004), Bik et al.~(2006) conclude
that the CO bandhead emission is produced in the warm (1500-4500 K),
dense, and optically thick region within a few AU from the central star.
The column densities required to model the CO line profiles 
in high-resolution spectra, $N(CO) \ge 10^{20} {\rm cm^{-2}}$ (Kraus et al.~2000,
Bik \& Thi 2004), suggest that CO molecules survive at these
radii as a consequence of self-shielding (Bik \& Thi 2004, and references therein).
Blum et al.~(2004) and Bik \& Thi (2004) showed that excellent fits
to the ro-vibrational CO 2-0 bandhead emission ($\lambda_0 = 2.2935\mu$m)
are obtained with rotating disk models, which particularly well 
account for the slope of the blue wing of the CO 
overtones, as well as the redshifted peak. Wind models 
from stellar or disk winds around young, massive objects, 
on the other hand, 
predict a vertical blue edge to the overtone emission band
and thus cannot account for the observed slow rise in the blue wing 
(see Kraus et al.~2000 for a detailed comparison of these 
scenarios). High-spectral resolution observations of 
the blue emission wing in the first CO overtone lent increasing
support to the rotating disk interpretation. In particular, 
Blum et al.~(2004) model the CO bandhead emission of 
four high-mass ($5-17\,M_\odot$) YSOs as Keplerian disks
originating in the inner 1 AU from the central star
with rotation velocities between 25 and 260 km/s. Spectral models
by Bik \& Thi (2004) reveal that the CO bandhead emission originates
in the inner parts of a thick accretion disk at radii 0.1-5 AU.


The source with the strongest CO bandhead emission (top spectrum 
in the right panel of Fig.~\ref{spectra}) displays a slanted ascent
in the blue wing of the first overtone. The slope in the blue wing
is particularly evident in comparison with the steep drop of the CO
absorption in the nearby background giant, where the envelope emission 
shows no signs of rotation. The shape of the blue wing is very similar
to the Keplerian disks modelled in Bik \& Thi (2004).
Although the moderate spectral resolution
prohibits the derivation of the rotational velocities of the Arches 
excess sources, all six detected bandheads are consistent with a 
slow increase in the blue wing.
We therefore conclude that the CO bandhead emission originates in the 
optically thick inner parts of rotating circumstellar disks. 

In addition to CO emission, two of the three excess source spectra display 
a weak Br$\gamma$ absorption feature. Circumstellar disks in young 
OB associations typically show strong Br$\gamma$ emission (e.g., 
Hanson et al.~1997, Bik et al.~2006).\footnote{Note that the more 
evolved Classical Be stars and B[e] supergiants also show near-infrared excess,
yet are characterized
by strong hydrogen emission lines. These objects are superluminous as
compared to their main sequence B-type counterparts, while the excess
sources in the Arches cluster are comparably faint.} 
In particular, the gas around
high-mass stars earlier than B3V is expected to be ionized by the 
EUV radiation from the central star. As indicated in the schematic model 
of a circumstellar disk around a massive protostar (Fig.~\ref{diskmodel}, 
adopted from Bik et al.~2005), Br$\gamma$ emission is expected from 
the inner disk rim as well as from the illuminated disk surface.
The spectral types of the disk-bearing stars with spectra are therefore 
likely later than B3V.

The increasing evidence of rotating disks around high-mass stars 
indicates that late O and early B stars form via disk accretion
in a similar fashion to their low-mass counterparts (where 
CO bandhead emission is also interpreted as evidence for 
thick accretion disks, e.g. Carr et al.~1993, Chandler et al.~1995),
which has a profound impact on the paradigm of high-mass star formation.

\subsection{Disk fraction and radial dependence}
\label{diskfracsec}

\subsubsection{Disk fraction}
\label{disksec}

With the conclusion that the excess sources in the Arches cluster are
dense circumstellar disks, and the membership information from proper motions,
we can estimate a robust disk fraction among
the B-star population of the Arches cluster. Given that all 19 $L'$-band
excess sources in the complete, $H < 19$ mag, sample with precise 
proper motion information are cluster members,
we have concluded in Sec.~\ref{lsec} on the basis of their exceptional colors 
that all 22 sources with $L'$-band excess and $H < 19$ mag
belong to the young Arches population. 
The main sequence population, 
on the other hand, is contaminated by foreground and background stars.
Hence, we only consider proper motion members in both the excess 
and main sequence samples to determine the cluster disk fraction.

The $H$-band completeness limit in the cluster core is the most
stringent limitation for the detection of main sequence stars.
The corresponding population of main sequence cluster members with 
$H < 19$ mag detected in all three filters is 
additionally limited by the $L'$-band detection limit. Completeness
limits of 13.5 - 14.5 mag imply that only early B-type stars with masses
in excess of $\sim 12\,M_\odot$ can be detected on the main sequence, where
excess emission is absent. The stellar mass range indicated for the 
$L'$-band excess sources is therefore largely below the $L'$ detection limit
of diskless main sequence cluster stars, such that the reference 
main sequence sample is incomplete.
The fraction of sources with $L'$-band excess (hereafter excess fraction),
defined as the ratio of sources with $L'$-band excess over all cluster
members with and without excess, 
$f_{exc} = N_{L-exc}/(N_{main seq} + N_{L-exc})$, would yield an upper
limit to the disk fraction. This limitation can be mitigated by not 
requiring main sequence stars to be detected in $L'$, which increases
the main sequence sample from 216 $HK'L'$ detections to a total
of 311 cluster members with $H < 19$ mag and $H-K' < 2.0$ mag.
The six excess sources falling into the main sequence $H-K'$ color regime
are excluded from the $H-K'$ main sequence sample. 
Five stars are substantially bluer than the Arches main 
sequence and may be foreground interlopers which incidentally have
similar proper motions. These stars with $H-K' < 1.2$ mag were also 
rejected. The remaining spread in color in the main sequence sample 
is caused by extinction variations. 
This selection leaves us with a main sequence reference sample of 300 sources. 

Excluding the three excess sources without 
(reliable) membership information, and the two sources beyond the 
$H$-band completeness limit in the core, $H=19$ mag, the comparison
sample of excess sources as defined in Sec.~\ref{lsec},
contains 19 stars, indicating a disk fraction
of $19/(300+19) = 6.0 \pm 1.6$\% among B-type stars in the Arches cluster.

\subsubsection{Radial distribution of excess sources}
\label{radsec}

As detailed in Sec.~\ref{intro}, the dense environment of a starburst cluster
is likely to destroy disks more rapidly than low-density
star-forming regions sparse in OB stars. This is particularly 
the case in the compact core of the cluster, where the 
exceptionally high density of more than
$10^5\,M_\odot {\rm pc^{-3}}$ could lead to significant 
destruction of disks in the immediate cluster center as 
compared to larger distances from the core.

With a total of 19 $L'$-band excess sources in the complete sample, 
we can begin to investigate the radial distribution of candidate disks in 
the Arches cluster. In Fig.~\ref{excess}, the histograms
of main sequence cluster members with $H < 19$ mag without excess 
(top solid line) 
and of members with $L'$-band excess (bottom solid line) are shown. 
Radial bins are chosen to contain similar total number counts
to minimise systematic uncertainties.
The relative fraction
of excess sources with respect to main sequence members
is labelled, along with the propagated uncertainty.
In addition, the absolute number counts of main sequence
members with and without excess are given. The right panel of 
Fig.~\ref{excess} displays the excess fractions vs.~radial
distance from the Arches cluster center. The substantial 
uncertainties shown as solid lines originate from the low
number of excess sources in each bin. The cluster center
distance corresponds to the median of main sequence stars 
in each bin; the radial range covered is indicated by 
dotted lines. Despite the large statistical uncertainties,
there is a strong indication that the disk fraction is 
significantly lower in the cluster core than at larger radii.

Before concluding that the excess fraction is depleted in 
the cluster core, incompleteness effects due to the higher 
stellar density have to be excluded as the source of the 
observed trend in the disk fraction. Artificial star simulations
were carried out in the magnitude range $16 < H < 21$ mag to probe
the completeness of the main sequence sample, and $12 < L < 16$ mag
to test the disk sample for incompleteness effects. A maximum of 50 
artificial stars were randomly inserted in the images to preserve 
the original stellar density. One-hundred such frames were created in 
each field, leading to a total of approximately 5000 artificial stars
in the probed magnitude ranges. The results of these tests are shown 
in Fig.~\ref{incfig}. The histograms (left) show the inserted (thin lines)
and recovered source counts (thick lines) in $L'$ and $H$. The asterisks
in the $L'$ histograms show the faintest disk source in each of the three
fields displayed. The right panel displays the recovery fractions vs. magnitude
for $L'$ and $H$. The completeness fractions are above 85\% in the outer fields
east1 and halo, and drop to 80\% in the cluster core in the {\sl faintest}
included magnitude bin, for $14 < L' < 15$ and $18 < H < 19$ mag. 
The completeness fractions are comparable on the main sequence and in the disk sample. 
Only very few sources are actually located in these faint magnitude bins; 
in the most crowding-limited cluster core sample, only 10 out of 208 main sequence 
sources are found 
in the faint bin, such that the 20\% correction adds a mere 2 sources to the core main 
sequence sample. Similarly, all disks in the cluster core are brighter than $L' = 14$ mag, 
although the simulation readily detects sources down to $L'=15$ mag. As crowding is 
not affecting detections in the outer fields, 
the incompleteness effects are entirely negligible at larger radii, as expected.
As a consequence, even when we account for incompleteness,
the disk fractions at all radii do not change appreciably. 

In conclusion, the disk fraction in the Arches cluster increases significantly
from 2.7\% in the cluster core, with $r < 0.16$ pc, to 9.7\% at $r > 0.3$ pc.

\section{Discussion}
\label{discsec}

\subsection{Disks in the Arches cluster}

The finding of disks in the Arches B-star population is unexpected
for two reasons. First, disk depletion depends on
the UV radiation of the host star. The $H$-band brightness of 
the disk-bearing stars suggests that the majority are B-type stars. 
As the characteristic UV evaporation timescale 
of a primordial disk around Herbig Be stars is less than 1 Myr
(Alonso-Albi et al.~2009), a disk lifetime of 2.5 Myr for 
B-type stars implies that the disks had to be massive initially.  
Secondly, disk destruction is expected to be accelerated in a
starburst cluster environment. 
The extreme UV radiation from numerous O-stars adds 
to the evaporation by the disk host star. At a density of 
$10^5\,M_\odot {\rm pc^{-3}}$, interactions cause additional
tidal truncation especially of the extended outer disks (Olczak et al.~2006), 
which in turn enhances the exposure of the inner disk to 
the cluster radiation field.
The detection of optically thick disks with substantial 
$L$-band excesses close to the cluster center suggests that 
the disk-bearing sources have migrated into the sphere of influence
of the central O-type stars only recently, or are viewed in projection 
toward the cluster center. 
The radial increase of the disk fraction toward larger cluster center
distances evidences the influence of the radiation field and high 
stellar density in the cluster core on primordial disks. 
Speculatively, one can envision a scenario wherein disk-bearing stars
migrate inward during the mass segregation process and once close 
to the cluster center lose their disks rapidly. 
Taking into account that some of the core sources might be in front 
of or behind the cluster center, the very low excess fraction of only 3\% 
in the cluster core is already an upper limit to the disk fraction inside
the core radius of 0.2 pc. The disk fraction of 10\% for stars with
$r > 7^{''}$ (0.28 pc) in the {\sl same stellar mass range} provides 
stringent evidence that the starburst cluster environment enhances 
the depletion of primordial disks as soon as their orbital motion 
moves disk-bearing stars close to the cluster center. 
In the future, the
increasing proper motion accuracy might provide trajectories of 
these sources to understand their orbital motion in the cluster.

From observations of lower-mass star-forming regions, we might expect
an increase in the disk fraction towards stars with later types, such as AFG stars.
If the same processes dominated disk destruction in a starburst cluster,
a substantial increase in the disk fraction in fields with deeper $L$-band
observations should be observed, which is not the case.
For instance, in the leading field with a peak in the luminosity function 
of $L'=15.5$ mag, the same number of 5 excess sources
is observed as in the shallower east1 and core fields with $L'$ peak
luminosities of only 13.5 and 14.0 mag. If the disk fraction would 
mostly be determined by the UV evaporation from the B-type host stars,
then the deeper detection limit in the leading field should yield 
a significantly larger fraction of disks than the shallower fields, 
especially when compared to the east1 field with similar stellar density.
The fact that an increase in the disk fraction towards fainter stars seems
not to be observed in the Arches cluster strengthens the interpretation that 
the starburst environment affects disk survival. 
It appears that stars later than B lose their disks more rapidly in 
a starburst cluster environment than in moderate star-forming regions,
indicating that the influence of the ambient UV radiation field and 
encounters indeed shorten the disk survival timescale. 

In Fig.~\ref{cmd_disks}, the Arches population is compared to disks 
from the most extensive Herbig Ae/Be survey of Hillenbrand et al.~(1992). 
The Herbig Be
star disks were shifted to the distance and foreground extinction of the 
Arches cluster. The Arches sources are systematically fainter than the 
primordial Group I star/disk systems in the Hillenbrand sample, which is the youngest
group with ages of a few $10^5$ Myr. In addition, the Arches disks cover $H-K'$ colors
closer to the main sequence than primordial Herbig Be disks, while still displaying 
significantly larger excesses than the Group III sources with depleted 
disks in the Hillenbrand sample. 
Also included in the CMD are $L$-band excess sources in NGC\,3603 at an age of 
1 Myr (Stolte et al.~2004, Harayama et al.~2008). The brightest
disks in NGC\,3603 are comparable in color and magnitude to the Arches sources. 
The difference between the disks in the Arches and NGC\,3603, compared 
to the primordial disks studied by Hillenbrand et al.~(1992), illustrates 
that the transition from primordial, optically thick dusty disks 
to evolved disks subject to evaporation and grain growth is a rapid process 
that happens at an age of approximately 1-2 Myr.

\subsection{Comparison with other young, massive clusters}
\label{compsec}

One of the best studied, nearby star-forming regions sufficiently massive
to host O-type stars is the Orion Nebula Cluster (ONC) with a disk fraction 
of 78\% in the predominantly low-mass population (Lada et al.~2000). 
The finding of significantly different disk fractions for high- and low-mass
stars (80\% for low-mass stars with $ M < 2\,M_\odot$, 42\% for OBA stars)
in this study corroborates the dependence of the disk depletion timescale
on the mass of the central star. The ONC provides a very early stage of a 
massive cluster at an age of 1 Myr, and allows predictions 
of the effects of close encounters with the UV radiating O-star.
Dynamical simulations of disk evaporation in the ONC
suggest that, while outer disks are depleted,
the inner disk is capable of surviving out to 10 AU in the immediate vicinity
of the central O6 star $\theta^1$C Ori (Scally \& Clarke 2001).
The possibility of survival of dense, inner disks around B-type stars 
up to ages of 10 Myr was also 
predicted theoretically from hydrodynamic simulations taking into account 
various levels of photoevaporation, disk-stellar wind interaction, mass loss
from disk winds, and the evaporative depletion by nearby O stars 
(Richling \& Yorke 2000, see Hollenbach et al.~2000 and Zinnecker \& Yorke 2007
for reviews).  

At an age of 2.5 Myr, the Arches cluster fills a rarely sampled 
gap between the very young UCH{\sc II}\ regions and deeply embedded 
clusters and the more evolved open clusters.
As one of the most massive young, compact star clusters in the 
Milky Way today, it is one of the rare loci where stars with masses 
above $100\,M_\odot$ were formed. With a population of at least 125 O-type 
stars (Sec.~\ref{intro}), the cluster UV radiation field of
$\sim 4\times 10^{51}$ photons s$^{-1}$  is intense (Lang et al.~2001).
When comparing the disk fraction in the Arches with moderate star-forming
regions, where disk fractions are studied among lower to intermediate
mass stars ($\sim 0.5-10\,M_\odot$), the major problem is to 
distinguish the effects of the mass dependence and the environment.

A comprehensive survey of Herbig Ae/Be stars in nearby OB associations
was carried out by Hern\'{a}ndez et al.~(2005). In the six star-forming 
regions with ages 3-16 Myr, these authors find that only a small fraction,
0 - 5\% of intermediate mass stars ($1.4 < M < 15\,M_\odot$),
are in the Herbig Ae/Be phase, which they 
interpret as the fraction of intermediate-mass stars having retained
dusty inner disks, in agreement with our findings above. From the comparison 
of disk frequencies around low ($M < 1\,M_\odot$) to intermediate mass stars, 
Hern\'{a}ndez et al.~(2005)
conclude that at an age of 3 Myr, the fraction of disks around intermediate
mass stars is ten times lower than around their low-mass counterparts.

In Fig.~\ref{diskage}, we reproduce the disk fraction vs.~age diagram 
for nearby star-forming regions from Haisch et al.~(2001), including recent 
results on a variety of clusters,
in addition to the Arches and NGC\,3603 data points. The cluster properties 
and literature references are summarized in Table \ref{clustab}.
At cluster ages similar to the Arches, 
the $\sigma$ Orionis cluster at an age of 3 Myr displays
a total thick disk fraction of $27 \pm 3$\% for all stars $> 0.1\,M_\odot$
(including some brown dwarfs), but this fraction drops to
$4 \pm 4$\% for stars with masses above $2\,M_\odot$ (Herbig Ae/Be,
Hern\'{a}ndez et al.~2007). 
Especially this latter fraction for intermediate-mass stars is 
strikingly similar to the Arches excess fraction.
The central 30 Doradus star-forming region with a mean 
age range of 2-3 Myr displays an $L$-band excess fraction of $42\pm 5$\%
(Maercker \& Burton 2005), with indications that even O-type to early B-type 
stars can retain their dense, inner disks for 2-3 Myr at least. The latter fraction 
is an upper limit to the disk fraction at 2-3 Myr, however, as the study 
by Maercker \& Burton (2005) covers the extended star-forming region 
surrounding the central cluster, and objects along the star-forming 
ridges with ages younger than 1 Myr are included in the $L$-band excess
fraction. Despite being an upper limit, the 30 Dor disk fraction is significanly
lower than expected
from the disk fraction-age relation derived in nearby star-forming regions 
(Fig.~\ref{diskage}).
As the $L$-band sample in 30 Dor is also dominated by early-type stars,
the effect of the cluster environment and the disk host stars cannot 
be distinguished in this region. 

At younger ages of 1-2 Myr, the massive star-forming regions
NGC\,3576, M17, and NGC\,3603 provide templates at earlier evolutionary stages.
In the more dispersed environment of NGC\,3576, Maercker et al.~(2006) find
a cluster disk fraction of $55 \pm 2$\% with a strong decrease from $95\pm 1$\%
in the immediate cluster core to $27 \pm 5$\% at a cluster center distance 
of 6 pc. It is noteworthy that Figuer\^{e}do et al.~(2002) analyze the 
early B/late O-type star \#48 in NGC\,3576, and conclude on the basis of CO
and H$_2$ emission, along with the absence of stellar features, 
that the source is most likely
a B1V star surrounded by a thick circumstellar disk or envelope, consistent 
with our interpretation of the Arches excess population. At an even younger age
of less than 1 Myr, the young, massive cluster NGC\,6618 at the center of the 
massive H{\sc II}\ region M17 has recently been investigated in detail with 
regard to its $JHKL$ population. Hoffmeister et al.~(2006) find an $L$-band excess
fraction of 41\% among a population of 201 M17 sources dominated by B-type stars. 
In medium-resolution 
spectroscopy, they find 9 stars with early B-type luminosities with CO emission 
features, and 7 or 78\% of these CO emitters feature $L$-band excesses.
Interestingly, two objects also show Br$\gamma$ and Pa$\delta$ in absorption. 
If we identify the M17 CO emission objects with the Arches $L'$-excess population,
and hence with B-type stars retaining optically thick gaseous disks, both the low 
fraction of $6 \pm 2$\% $L'$-band excess sources and the detection
of CO emission in all three of our excess sources with spectra is consistent with 
a more evolved counterpart of the younger M17 population. The denser and more
hostile environment of the Arches cluster additionally causes the disk fraction 
to be substantially depleted as compared to the younger, and less dense,
environment of M17.

At an age of 1 Myr, the compact, massive cluster NGC\,3603\,YC 
is most similar to the Arches cluster in its core density and total mass
(see, e.g., Stolte et al.~2006 for a detailed comparison of the cluster properties).
A disk fraction of $27\pm 3$\% was measured in the central 1 pc for stars in the 
mass range $1 < M < 20\,M_\odot$ (Stolte et al.~2004),
and a lower fraction of 12\% $L$-band excess sources was observed among the main sequence
OB star population in the radial range from 0.2 to 0.8 pc from seeing-limited $JHKL$ 
observations. Using high-spatial resolution imaging,
Harayama et al.~(2008) recently resolved the central 0.4 pc of the
NGC\,3603\,YC cluster core, and derived an $L$-band excess fraction of $24\pm 10$\% 
for the entire population with $M > 1\,M_\odot$, 
and $22\pm 10$\% for main sequence stars with $M > 4\,M_\odot$ only. 
The higher excess fraction in the cluster core is most likely a consequence
of resolving the fainter excess population in the high-angular resolution 
$L$-band observations. This disk fraction is substantially lower than the 
characteristic disk fraction of nearby 1-2 Myr star-forming regions (Fig.~\ref{diskage}).
In the disk-fraction vs.~age diagram, the three densest clusters Arches, NGC\,3603,
and NGC\,6618 (M17) display the lowest disk fractions in their respective age group.
This is not an effect of the starburst cluster environment exclusively, as the disk
fractions of Herbig Ae/Be stars in IC 348 and $\sigma$ Ori are comparably low,
suggesting that disk depletion is dominated by the stellar mass of the 
host star.
The low disk fractions observed in all of these massive clusters among early-type 
stars is consistent with a more rapid depletion of disks around high-mass stars
than lower-mass cluster members. There is increasing evidence that young, massive
clusters comprised of a large population of OB-type stars have lower disk fractions
than clusters dominated by low-mass stars. In Fig.~\ref{diskage},
the massive clusters Arches, NGC\,3603, 30 Dor, NGC\,3576, NGC 6618, as well as 
$\sigma$ Orionis all fall below the linear trend derived by Haisch et al.~(2001). 
It is particularly striking that the four massive clusters where predominantly 
B-type stars are probed (including the Herbig Ae/Be data point for $\sigma$ Ori),
appear to follow the same fractional decrease in the disk fraction with age 
shifted to lower total disk fractions. 

We thus conclude that the $L'$-band excess of the Arches sources
is produced by remnants of massive accretion disks around mid to late B-type stars 
in a more evolved state than the CO and Br$\gamma$ emission disks observed in
the younger UCH{\sc II}\ regions and star-forming clusters such as the Trapezium, 
M17, NGC\,3576, NGC\,2024 with ages $\le 1$ Myr.

\section{Summary}
\label{sumsec}

We report the detection of $L'$-band excess sources in the 
Galactic center Arches cluster. Out of 24 excess
sources with $K'-L' > 2.0$ mag, all 21 sources with precise proper 
motion measurements are proper motion members of the cluster.
The CO bandhead emission detected in three sources with K-band
spectroscopy identifies these sources as optically thick disks. 
The comparison with disk simulations in similarly young star-forming 
regions suggests that the CO emission and $L'$-band excess arise in the 
inner few AU from the central star. 
The $H$-band magnitude range of these sources indicates that
the sample consists of Herbig Be stars, possibly with a small contribution 
of early Herbig Ae stars. 

From the complete sample of proper motion members with $H < 19$ mag, 
we derive a disk fraction of $6 \pm 2$\% among B-type stars
in the central Arches cluster.
At the young cluster age of 2.5 Myr, this disk fraction is significantly 
lower than expected from the disk fraction-age relation found 
in star-forming regions with lower central density at ages 2-3 Myr. Even 
when compared to B-star samples in other massive star-forming
clusters at similar ages, such as 30 Dor or NGC\,2264, 
the dense Arches and $\sigma$ Ori clusters feature the lowest disk fraction 
measured at this age. The disk fraction in the Arches cluster increases 
from $2.7 \pm 1.8$\% in the cluster core ($r < 0.16$ pc)
to $5.4 \pm 2.6$\% at intermediate radii ($0.16 < r < 0.3$ pc)
and $9.7 \pm 3.7$\% outside the cluster center ($r > 0.3$ pc). 
The preferential depletion of disks in the cluster 
core evidences disk destruction by UV radiation, winds, and tidal encounters 
of massive stars. The comparison with the younger starburst cluster NGC\,3603
and more moderate star-forming regions provides a strong observational
indication that disks are more rapidly depleted in a starburst environment.

In the case of a very dense cluster such as the Arches 
the detection of a significant population of circumstellar
disks nevertheless comes as a surprise. In the presence of on the 
order of 125 O-type
stars contributing to the intense cluster radiation field, which 
ionizes the surrounding molecular clouds (Lang et al.~2001),
disk destruction must be very efficient. The survival of 
disks over several Myr in such a dense cluster environment 
sheds new light on the emergence of debris disks around high-mass
stars, and the possibility of forming, and detecting, planetary 
systems in dense, massive clusters.

Combining the lower total disk fraction 
observed in dense, massive clusters with the depletion of disks in the core 
of the Arches cluster as compared to larger radii (Sec.~\ref{radsec}),
provides strong indication that clusters with substantial OB star 
populations have lower disk fractions especially in the immediate environment
of high-mass stars. These results
suggest that the rate of disk depletion and thus the disk lifetime depend not only 
on the spectral type of the disk host stars, but also on the total 
cluster mass and density as an indicator for the ambient UV radiation field 
and the frequency of dynamical interactions with the massive OB stars 
specifically in the cluster cores.

\acknowledgements

This work would not have been possible without the intense effort 
and dedication of the Keck LGS-AO staff. We are deeply grateful
for their support enabling these observations. The W. M. Keck Observatory
is operated as a scientific partnership among the California Institute
of Technology, the University of California, and the National Aeronautics
and Space Administraction. The Observatory was made possible by the generous
financial support of the W. M. Keck Foundation. The authors wish to recognize
and acknowledge the very significant cultural role and reverence that the 
summit of Mauna Kea has always had within the indigenous Hawaiian community.
We are most fortunate to have the opportunity to conduct observations from 
this mountain.
AS acknowledges support by the German Science Foundation Emmy Noether
programme (STO 496/3-1). 
This work was supported by NSF grant AST 04-06816 and by the
Science and Technology Center for Adaptive Optics,
managed by the University of California at Santa Cruz under
cooperative agreement AST 98-76783.

Facilities: Keck:II (NIRC2), VLT (NAOS-CONICA)

\appendix
\section{Galactic center $L'$ calibration sources}

The absolute zeropoint of the Arches core field $L'$ observations was 
derived by observing the central $10^{''}$ (0.4 pc) around Sgr A\* during the same 
night as the cluster center.
Detailed descriptions of the observing strategy can be found in 
Ghez et al.~(2005), Hornstein et al.~(2007), and Lu et al.~(2009).
Table A.1 lists the non-variable GC sources used as calibrators 
for the absolute $L'$ zeropoint in the GC. These sources are identified in 
Fig.~A.1 (see also Lu (2009) for the naming convention).
The zeropoint calibration is detailed in Sec.~\ref{lzero}.
The foreground extinction to the central parsec, $A_V \sim 26-36$ mag, is
comparable to the foreground extinction toward the Arches.
The advantage of this central reference field is that there are numerous 
calibration sources with comparably red colors as the Arches cluster stars.

Fig.~\ref{zptplots} shows the zeropoint derived from the combined image of
19 frames taken in the Galactic center observing sequence. The combined image
yields a zeropoint of $ZPT_L' = 23.196 \pm 0.096$ mag, while the individual frames 
yield a median zeropoint of $ZPT_L' = 23.417 \pm 0.130$ mag. The zeropoint variations 
among the Arches individual images is significantly smaller with a standard 
deviation of only 0.039 mag, indicating that atmospheric conditions had equilibrated
when the cluster was observed. Hence, a mean zeropoint between the combined Galactic 
center image and the individual frames of 23.300 was applied to the combined Arches
core image. The zeropoint uncertainty is estimated to be $\sim 0.10$ mag due to 
atmospheric/temperature variations during the observing sequence.

\begin{table}[h4]
\tablenum{A.1}
\caption{\label{gctab} $L'$ calibration sources in the Galactic center.}
\footnotesize
\begin{tabular}{lrrrrrrrrr}
\hline
  Name    &    Radius [``] & $\delta\,$RA & $\delta\,$DEC & $H$ ref$^a$  &   $K'$ ref$^a$ &  $L'$ ref$^a$  &  $H$ cal$^b$  & $K'$ cal$^b$ &  $L'$ cal$^b$ \\ 
\hline
  IRS16C  &    1.209 &    1.112 &    0.476 &  11.940 &   9.792 &   8.090 &  11.945 &   9.952 &   8.182 \\  
  IRS16NW &    1.178 &    0.000 &    1.178 &  12.080 &  10.061 &   8.452 &  12.080 &  10.197 &   8.445 \\
  IRS33E  &    3.121 &    0.673 &   -3.047 &  12.260 &  10.948 &   8.545 &  12.136 &  10.269 &   8.492 \\
  IRS29S  &    2.028 &   -1.818 &    0.901 &  13.310 &  11.262 &   9.710 &  13.579 &  11.338 &   9.600 \\
  S1-23   &    1.701 &   -0.896 &   -1.446 &  13.980 &  11.854 &  10.003 &  13.861 &  11.743 &   9.961 \\
  S3-2    &    3.036 &    2.981 &    0.575 &  14.060 &  12.027 &  10.480 &  13.874 &  12.111 &  10.528 \\
  S2-8    &    2.109 &   -1.949 &    0.806 &  13.770 &  12.148 &  11.070 &  13.971 &  12.223 &  10.886 \\
  S3-5    &    3.106 &    2.896 &   -1.123 &  14.430 &  12.140 &  10.132 &  14.246 &  12.032 &  10.024 \\
  S3-10   &    3.471 &    3.299 &   -1.080 &  13.610 &  12.327 &  10.650 &  13.699 &  12.116 &  10.725 \\
  S3-19   &    3.169 &   -1.557 &   -2.760 &  13.890 &  12.470 &  10.560 &  13.863 &  12.018 &  10.517 \\
  S1-20   &    1.615 &    0.372 &    1.571 &  15.000 &  12.596 &  10.822 &  15.102 &  12.765 &  10.829 \\
  S1-22   &    1.686 &   -1.608 &   -0.508 &  14.650 &  12.693 &  10.947 &  14.657 &  12.646 &  10.868 \\
  S4-129  &    4.224 &    3.633 &   -2.155 & \nodata &  12.744 &  10.420 &  14.468 &  12.208 &  10.489 \\
  S1-5    &    0.979 &    0.368 &   -0.907 &  14.930 &  12.727 &  10.918 &  14.771 &  12.640 &  10.928 \\
  S3-30   &    3.343 &    1.690 &   -2.884 &  14.540 &  12.992 &  10.837 &  14.404 &  12.433 &  10.838 \\
  S4-6    &    4.186 &    3.245 &   -2.644 &  15.580 &  13.370 &  11.165 &  15.050 &  12.852 &  11.225 \\
  S3-6    &    3.165 &    3.164 &    0.080 &  14.890 &  12.827 &  11.075 &  14.991 &  12.818 &  11.130 \\
  S2-22   &    2.294 &    2.282 &   -0.228 &  14.740 &  12.881 &  10.882 &  14.641 &  12.888 &  11.245 \\
  S1-34   &    1.273 &    0.848 &   -0.949 &  14.770 &  13.013 &  11.773 &  14.515 &  13.069 &  11.503 \\
  S1-1    &    0.965 &    0.965 &    0.033 &  14.930 &  13.052 &  11.730 &  14.843 &  13.130 &  11.692 \\
  S2-5    &    2.022 &    1.860 &   -0.793 &  15.170 &  13.231 &  11.770 &  15.250 &  13.312 &  11.900 \\
  S0-13   &    0.686 &    0.537 &   -0.427 &  15.430 &  13.408 &  11.765 &  15.375 &  13.407 &  11.861 \\
  S1-25   &    1.733 &    1.625 &   -0.600 &  15.470 &  13.466 &  11.785 &  15.409 &  13.366 &  11.815 \\
  S2-21   &    2.322 &   -1.656 &   -1.627 &  15.380 &  13.526 &  11.927 &  15.394 &  13.445 &  11.750 \\
  S0-14   &    0.807 &   -0.756 &   -0.282 &  15.520 &  13.656 &  12.360 &  15.662 &  13.763 &  12.292 \\
  S0-15   &    0.946 &   -0.906 &    0.272 &  15.620 &  13.660 &  12.183 &  15.678 &  13.732 &  12.198 \\
  S2-77   &    2.754 &   -1.705 &   -2.163 &  15.420 &  13.857 &  11.757 &  15.227 &  13.603 &  11.589 \\
  S3-29   &    3.332 &    1.442 &   -3.004 &  15.480 &  14.079 &  12.225 &  15.571 &  13.628 &  12.228 \\
  S2-25   &    2.508 &    0.763 &   -2.389 &  15.880 &  14.106 &  12.175 &  15.952 &  13.820 &  12.189 \\
  S2-58   &    2.417 &    2.135 &   -1.133 &  16.000 &  14.079 &  12.418 &  16.013 &  14.084 &  12.597 \\
  S2-2    &    2.060 &   -0.569 &    1.980 &  15.460 &  14.090 &  12.860 &  15.828 &  14.052 &  12.868 \\
  S1-13   &    1.392 &   -1.069 &   -0.891 &  16.380 &  14.180 &  12.247 &  16.126 &  14.050 &  12.350 \\
  S3-8    &    3.388 &    3.362 &   -0.421 &  15.630 &  14.014 &  12.180 &  15.654 &  13.874 &  12.292 \\
  S0-9    &    0.554 &    0.143 &   -0.535 &  16.150 &  14.348 &  12.837 &  16.002 &  14.173 &  12.652 \\
  S3-22   &    3.157 &   -0.316 &   -3.142 &  13.260 &  11.671 &   9.480 &  13.133 &  11.167 &   9.475 \\
  S4-3    &    4.104 &    4.100 &    0.172 &  14.870 &  13.223 &  11.435 &  14.874 &  13.249 &  11.481 \\
  S3-178  &    3.410 &   -0.395 &   -3.387 &  15.220 &  13.467 &  11.385 &  15.098 &  13.067 &  11.356 \\
  S4-169  &    4.360 &    4.350 &    0.299 &  15.230 &  13.674 &  12.260 &  15.313 &  13.881 &  12.228 \\
  S4-1    &    3.987 &    3.976 &   -0.299 &  15.400 &  13.173 &  11.737 &  15.442 &  13.376 &  11.737 \\
  S4-161  &    4.338 &    4.333 &   -0.207 &  15.840 &  13.897 &  11.907 &  15.893 &  13.926 &  11.997 \\
  S3-134  &    3.272 &   -1.329 &   -2.991 &  16.080 &  13.807 &  12.187 &  16.045 &  13.848 &  12.074 \\
  S3-88   &    3.095 &   -0.802 &   -2.989 &  16.480 &  14.466 &  12.653 &  16.056 &  14.324 &  12.665 \\
  S4-4    &    4.216 &    3.560 &   -2.260 & \nodata &  12.793 &  10.410 &  13.937 &  11.965 &  10.323 \\
  S3-13   &    3.843 &    3.729 &    0.927 & \nodata &  13.578 &  11.847 &  15.855 &  13.566 &  11.940 \\
\hline
\end{tabular}
\footnotesize{$^a$ Columns 5-7 refer to reference magnitudes from earlier NGSAO observations in the central parsec,
calibrated against photometry in Cl\'{e}net et al.~(2001, 2004) and Blum et al.~(1996).}
\footnotesize{$^b$ Columns 8-10 refer to photometry taken on 2006 May 21, and calibrated with respect to the reference
magnitudes presented in columns 5-7. A comparison of $L'$ calibrated and reference magnitudes is shown in Fig.~\ref{zptplots}.}
\end{table}

\begin{figure}
\figurenum{A.1}
\includegraphics[width=16cm]{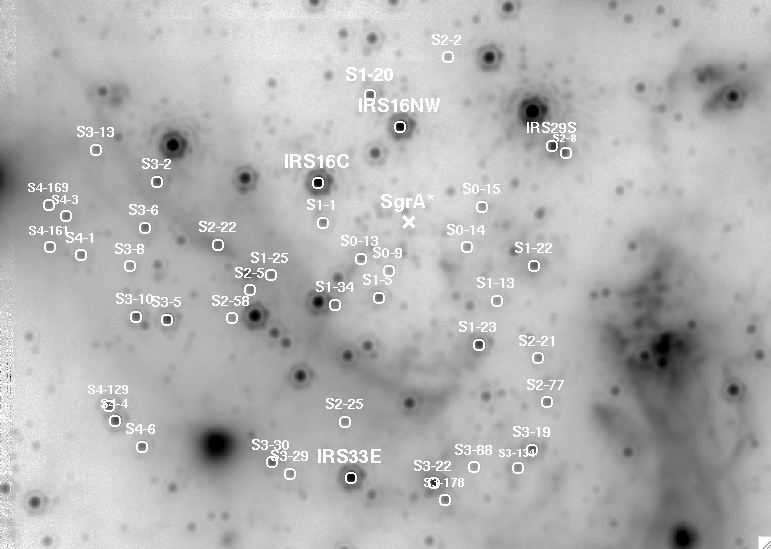}
\caption{\label{gc_calib_sources} 
Finding chart of non-variable $L'$ calibration sources located 
in the Galactic center (North is up, East is to the left).}
\end{figure}

\begin{figure}
\figurenum{A.2}
\includegraphics[width=6cm,angle=90]{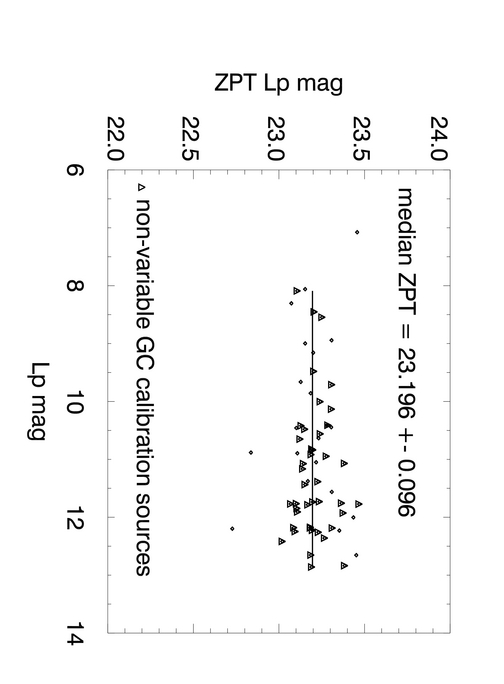} \\
\includegraphics[width=6cm,angle=90]{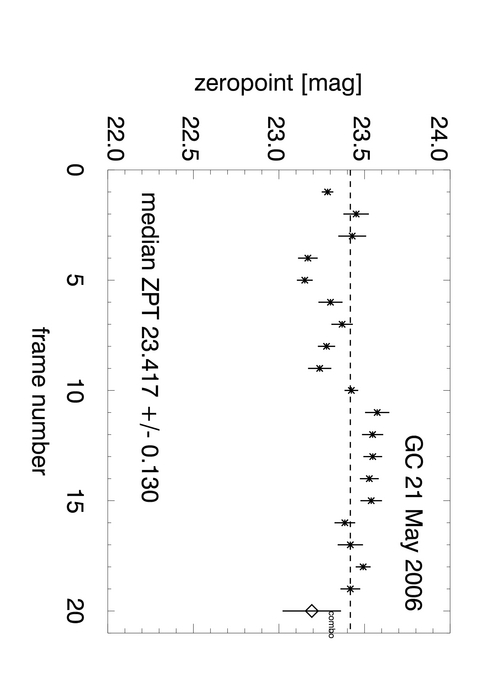} \\
\includegraphics[width=6cm,angle=90]{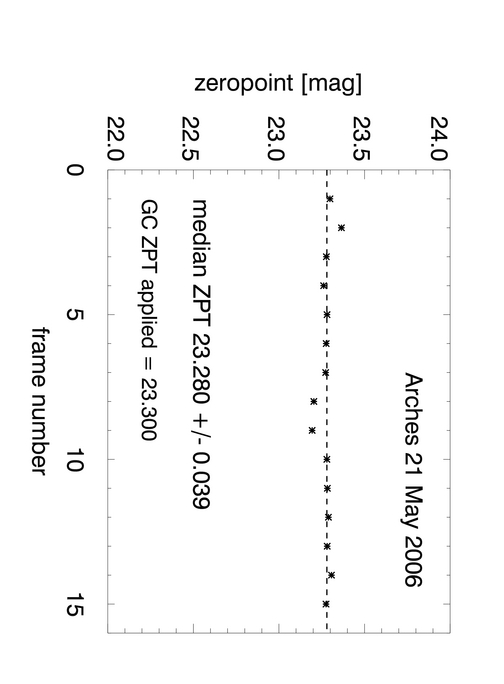} \\
\caption{\label{zptplots}
Zeropoint determination of GC sources and variations throughout the observing sequence.
{\it Upper panel:} $L'$ zeropoint of non-variable sources in the Galactic center, derived from 
reference magnitudes provided in Table \ref{gctab}. This zeropoint was derived from the combined
image composed of 19 frames. {\it Middle panel:} Zeropoint variation between individual Galactic center
frames from the Galactic center $L'$ observing sequence. {\it Bottom panel:} 
Zeropoint variations between individual Arches frames as observed on the core field,
after a GC zeropoint of 23.300 has been applied.}
\end{figure}

\section{Arches $L'$ zeropoint calibration and $L$-band extinction notes}

The Arches $L$-band observations were obtained with the 
Mauna Kea $L'$ filter. While the GC high-resolution 
observations by Cl\'{e}net et al.~(2001) were obtained with 
the similar VLT $L'$ filter 
($\lambda_c = 3.8 {\rm \mu m}$, $\Delta\lambda=0.62 {\rm \mu m}$),
earlier GC observations from Blum et al.~(1996)
used as calibration reference also by Cl\'{e}net et al.~(2001)
were observed with the obsolete $L$ filter, which has a comparable
width but lower central wavelength than $L'$
($\lambda_c = 3.45 {\rm \mu m}$, DePoy \& Sharp 1991, 
see also Simons \& Tokunaga (2001) for filter definitions).
Earlier extinction laws towards the central galaxy by 
Rieke \&  Lebofsky (1985) and Mathis (1990) were determined 
from the obsolete $L$ filter, while newer determinations 
make use of the Spitzer database, employing the IRAC 
$\lambda_c = 3.6 {\rm \mu m}$ band. Extrapolation of the 
standard extinction law reviewed in Mathis (1990) with a 
slope of $\alpha = 1.70$ yields $A_L/A_V = 0.058$ vs. 
$A_{L'}/A_V = 0.043$. The difference for the $K'$-band, on 
the other hand, is with $A_K/A_V = 0.112$ vs. $A_{K'}/A_V = 0.114$
negligible, and the $H$-band central wavelength and bandwidth
did not change significantly.

$A_V$ values of the Arches $L'$-excess sources span a 
range from 23 to 28 mag with an average of 25.3 mag (Table \ref{hkltab}).
These extinctions were derived from the four nearest 
cluster members to each excess source, and are representative
of the cluster population. With the above values for $L$-band 
extinction, we expect stars with $A_V\sim 23$ mag to be brighter in 
$L'$ by $\Delta(L-L') = (0.058 - 0.043) * A_V = 0.34$ mag 
with respect to $L$, and sources with 
$A_V\sim 28$ mag brighter in $L'$ by 0.42 mag. The mean offset 
between the $L$ and $L'$ filters for the median extinction 
of the GC calibration sources, $\Delta (L-L') = 0.50$ for 
$A_{V,GC} = 33.1$ mag, is corrected by calibrating 
the Cl\'{e}net et al.~(2001) $L'$ measurement with Blum et al.~(1996)
$L$-band magnitudes. The residual, extinction dependent offset
expected for Arches cluster members with 
$A_{V,Arches} = 25.3$ mag and $\Delta(L-L') = 0.38$ mag
is therefore 
$\Delta(L-L')_{GC} - \Delta(L-L')_{Arches} = 0.50 - 0.38 = 0.12$ mag,
which explains the offset of Arches cluster members from the 
standard extinction law in the two-color diagram, as discussed 
in Sec.~\ref{lzero}. The range in extinctions in the cluster 
population $22.7 < A_V < 27.9$ mag corresponds to $0.43-0.34 = 0.08$ mag, 
which is within our uncertainties of $\sigma_{L'} \sim 0.1$ mag. 
This offset would appear
as a bend in the reddening track in the two-color diagram 
(see Stead \& Hoare 2009 for a discussion on non-straight
reddening tracks). This bend is not detectable given the 
photometric uncertainties discussed in Sec.~\ref{photsec}.

Note that the $L$-band calibration from GC sources, where 
the bulk offset between $L$ and $L'$ was corrected, implies
that the standard $L$-band extinction law represents the proper 
extinction value of $A_L / A_V = 0.058$, which has been used  
accordingly. The value of $A_{L'} / A_V = 0.043$ extrapolated
from the central wavelength of the $L'$ filter yields a reddening
track that is too flat given this $L$-band calibration procedure.

\setcounter{table}{0}


\begin{figure*}
\includegraphics[width=16cm]{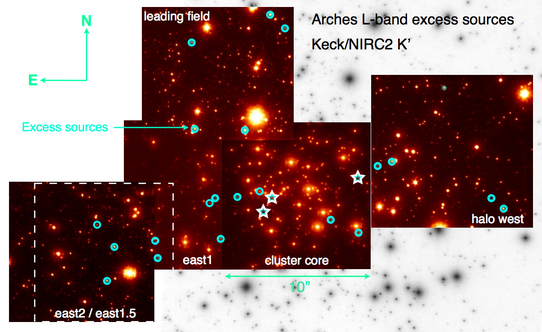}
\caption{\label{mosaic} Arches Keck/NIRC2 K' mosaic image,
with North up and East to the left. The tip-tilt reference
star is indicated by the large cicle, and $L$-band excess sources are 
encircled in light blue.
The cluster core is visible as the area of high stellar density in the center 
of the image,
fields east1 and east2 are offset to the East and Southeast, 
the leading field is oriented along the proper motion axis to the North,
and the halo field is located to the West and off the motion axis.
The intermediate field east1.5 is marked with a dashed box. Proper 
motion coverage is provided in the total area of the underlying grey-scale 
VLT-NAOS/CONICA image and the field east1.5.
The three excess sources with SINFONI spectra are labelled with star symbols.}
\end{figure*}


\begin{figure*}
\includegraphics[width=16cm]{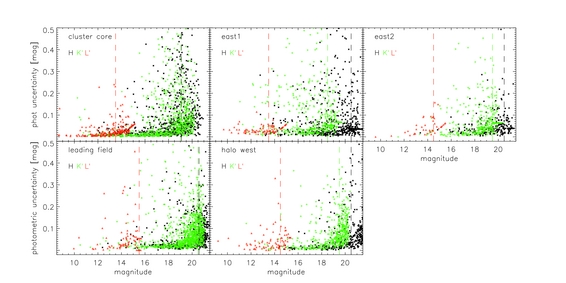}
\caption{\label{photuncer} Photometric uncertainties for each field 
({\sl black: H, green: K', red: L'}). The dashed lines indicate the observed peaks
of the luminosity functions (LF). In the cluster core and leading field, the H and K 
LF peak is identical at 19 mag and 20.5 mag, respectively.} 
\end{figure*}


\begin{figure*}
\includegraphics[width=16cm,angle=0]{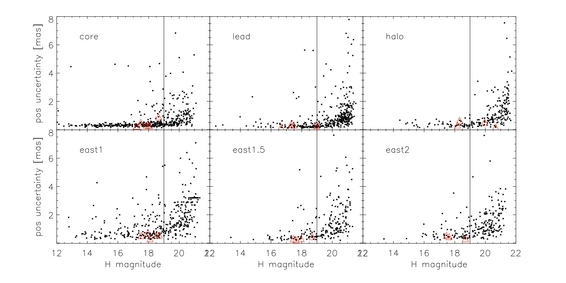}
\caption{\label{posuncer} Positional uncertainties for each field
measured from NIRC2 $K'$ observations. The solid lines mark the imposed
completeness limits at $H=19$ mag. $L'$-band excess sources are 
shown as triangles (red).}
\end{figure*}


\begin{figure*}
\includegraphics[width=12cm, angle=90]{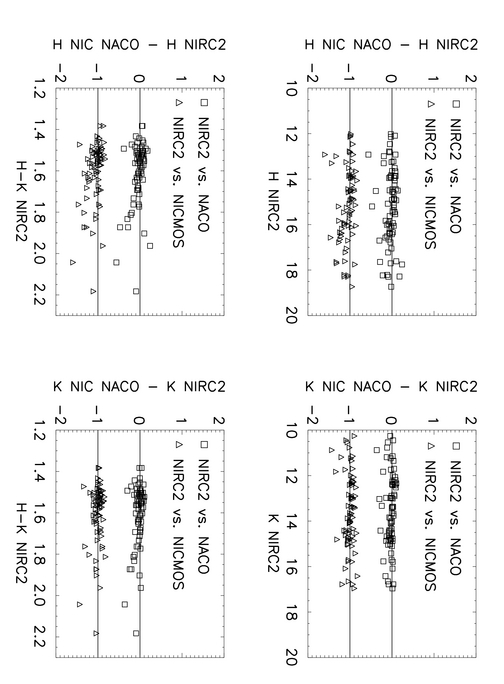}
\caption{\label{colorterms} Color terms beetween NIRC2 $H$ and $K'$ observations
and NACO $H$ and $K_s$ as well as NICMOS $F160W$ and $F205W$ calibrations.
The {\sl top panels} show that there is no systematic trend between calibrated and reference 
magnitudes in the covered brightness regime of calibration sources. 
The residual magnitude differences vs.~calibrated $H-K'$ color ({\sl bottom panels}) display 
no significant color terms. NIRC2 vs. NICMOS comparisons are shifted by -1 for clarity.}
\end{figure*}


\begin{figure*}
\includegraphics[width=12.4cm,angle=90]{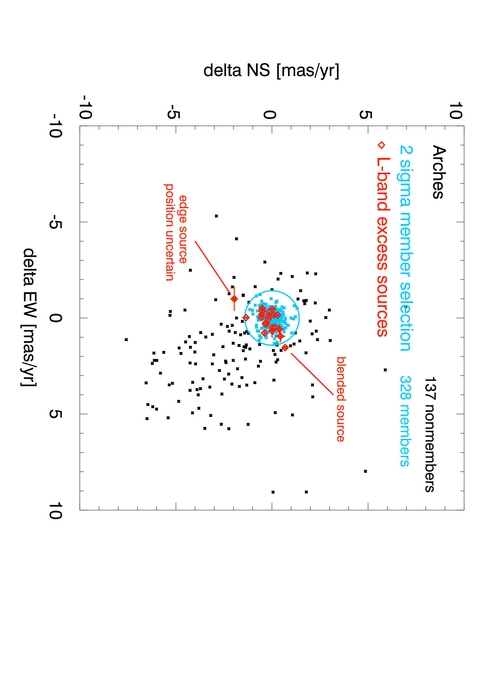}
\caption{\label{xymove} Proper motion membership selection of stars
with $H < 19$ mag.
Cluster members are selected within 2 sigma of the FWHM
derived from a gaussian fit to the motion distribution.
Excess sources are marked as red diamonds. Positional uncertainties for 
most excess sources are smaller than the symbol size. The large uncertainty
of the one excess source below the proper motion selected cluster member sample
(cyan) does not permit membership determination for this source. The second 
formal non-member in the excess sample is blended with a nearby star in the 
NACO image. Both sources are excluded from the member disk sample.}
\end{figure*}


\begin{figure*}
\hspace*{-2cm}
\includegraphics[width=7.4cm,angle=90]{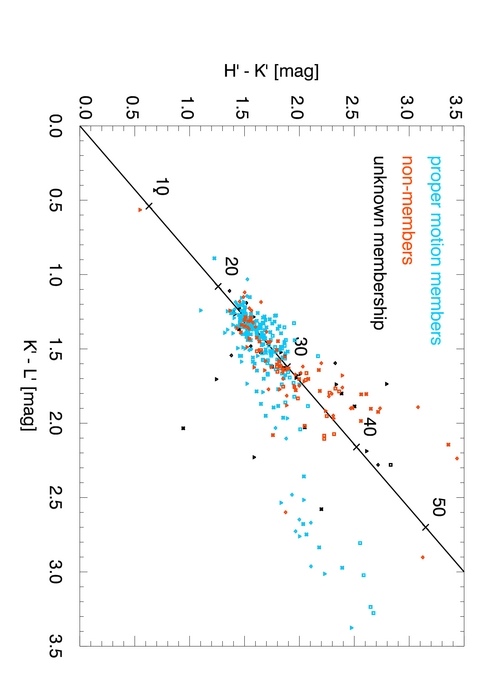}
\includegraphics[width=7.4cm,angle=90]{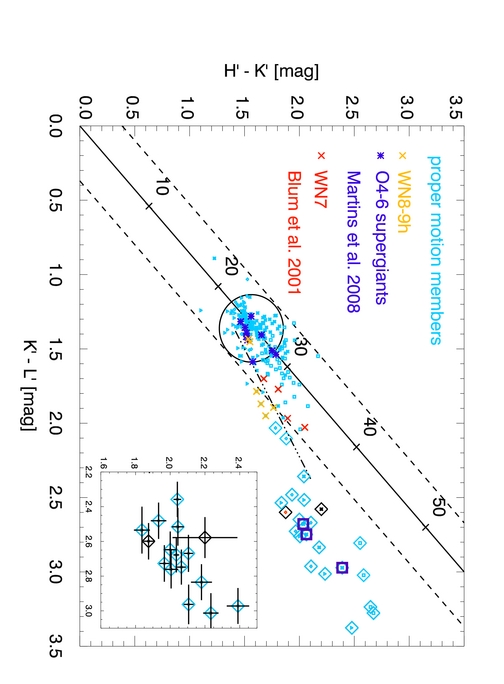}
\caption{\label{ccd_hkl} $H-K'$, $K'-L'$ two-color diagram of sources
in the Arches cluster fields. The solid line indicates
the reddening path of an A0 star using the extinction law as measured towards 
the GC (Rieke \& Lebofsky 1985).
{\sl Left:} Proper motion members (light blue)
preferentially cluster below extinctions of $A_V < 30$ mag, 
while non-members (red) scatter to extinction values as high as 
$A_V \approx 50$ mag, as expected along the GC line of sight.
Sources with $L'$-band excess are located significantly to the 
right of the reddening vector. 
{\sl Right:}  $H-K'$, $K'-L'$ diagram with Arches members.
The two excess sources without membership information (black diamonds)
are included for completeness. 
The ellipse marks the 2-sigma standard deviation in $H-K'$ 
and $K'-L'$ (note the different axis stretch in $H-K'$ and $K'-L'$,
causing the ellipse to appear as a circle), and the dashed lines
are the tangents of the uncertainty ellipse parallel to the reddening 
vector.
Main sequence cluster members with enhanced foreground extinction 
are expected to occupy the space between the dashed lines.
Excess sources are selected redwards of the lower dashed line if 
their photometric uncertainty indicates a significant offset 
from this line.
The excess sources (diamonds) stand out at colors redder than 
$K'-L' = 2.0$ mag. The inset shows excess sources 
with photometric uncertainties. The classical T Tauri locus 
is included as a dash-dotted line for reference (CTTS, Meyer et al.~1997).
Blue asterisks and yellow crosses denote young, high-mass stars with
known spectral types (Martins et al.~2008). Red crosses are evolved 
bright clusters members ($K' < 11.5$ mag) without spectral types.
The bulk of the excess
source population has substantially more excess than expected for 
young, lower-mass stars with disks (CTTS), or evolving hydrogen-rich 
Wolf-Rayet stars (yellow crosses) surrounded by dusty envelopes.}
\end{figure*}


\begin{figure*}
\hspace*{-1.7cm}
\includegraphics[width=7.4cm,angle=90]{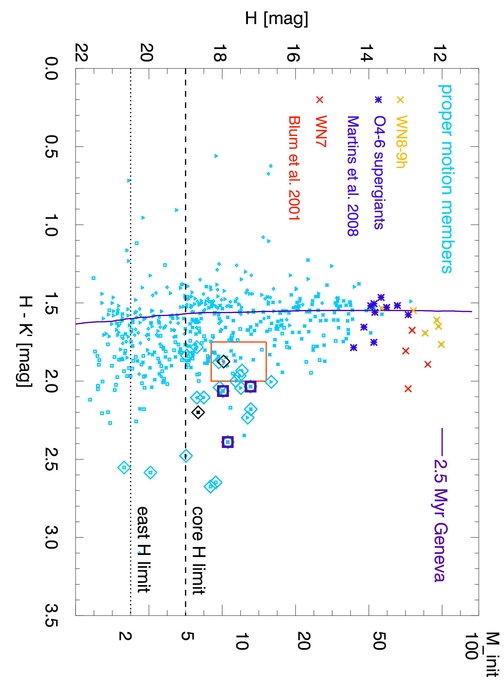}
\includegraphics[width=7.4cm,angle=90]{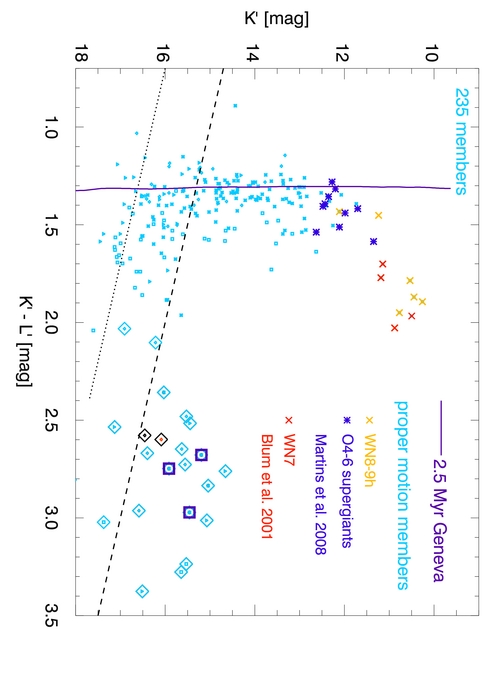}
\caption{\label{cmd} Keck/NIRC2 color-magnitude diagrams of Arches 
proper motion members.
$K'-L'$ excess sources are labeled (diamond symbols). The dashed lines 
indicate the completeness limit due to the turnover in the $H$-band LF
(left panel) and the $L'$-band LF (right panel)
in the crowding-limited core and the less dense east fields.
A 2.5 Myr Geneva isochrone with solar metallicity, shifted to an 
adopted distance to the GC of 8 kpc (Ghez et al.~2008)
and $A_V = 26$ mag (Stolte et al.~2005), using the Rieke \& Lebofsky (1985)
extinction law measured towards the GC, is shown for reference.
The location of red clump stars in the inner bulge is indicated 
by the red box. 
The three sources with SINFONI spectroscopy are marked by blue boxes. 
Left: The $H-K'$ vs.~$H$ CMD reveals that some $L'$-band excess
sources blend with the exincted population along the GC line of sight
towards the Arches, while several excess sources display significant 
$K'$-band excess as well. Right: In the $K'-L'$ vs.~$K'$ CMD, $L'$-band excess sources
are clearly distinct, displaying offsets of $\ge 0.5$ mag from main sequence stars. 
While contamination with the main sequence and possible red clump interlopers at 
the GC would be a major problem in $H-K'$ without membership information, 
the ambiguity is resolved in $K'-L'$.}
\end{figure*}


\begin{figure*}
\hspace*{-1.7cm}
\includegraphics[width=7.4cm,angle=90]{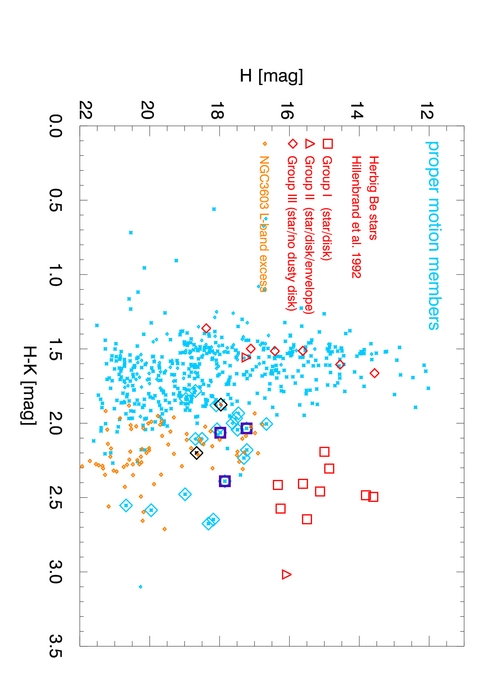}
\includegraphics[width=7.4cm,angle=90]{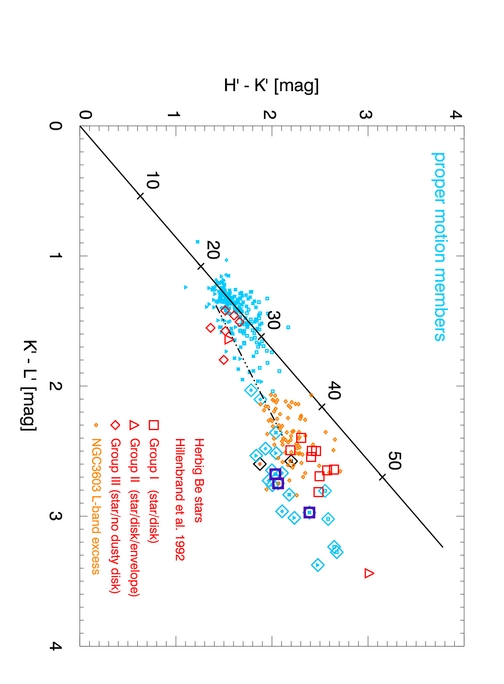}
\caption{\label{cmd_disks} Same as Fig.~\ref{cmd} and Fig.~\ref{ccd_hkl}
with Herbig Be disks from Hillenbrand et al.~(1992) and NGC\,3603 
$L'$ excess sources from Stolte et al.~(2004) included.
The Herbig Be and NGC\,3603 sources were adjusted to the distance 
and foreground extinction of the Arches cluster. 
Herbig Be Group I star/disk systems with ages $< 1$ Myr occupy the same $K'-L'$ 
color regime as the Arches excess sources, but are systematically brighter.
The Arches disks at an age of 2.5 Myr may be similar star/disk systems 
at a later evolutionary state.}
\end{figure*}


\begin{figure*}
\includegraphics[width=8cm]{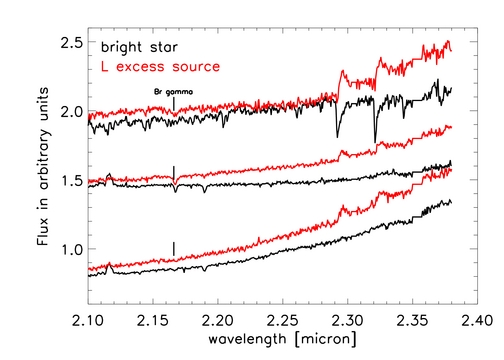}
\includegraphics[width=8cm]{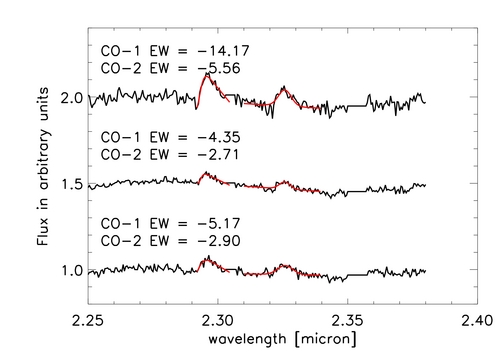}
\caption{\label{spectra} VLT/SINFONI $K$-band spectra of three of the Arches $L'$
excess sources. The black lines are spectra of the nearest bright stellar 
neighbour, while the red lines are spectra of the excess sources. While the  
spectra of neighbouring bright sources show stellar emission and absorption lines, 
the excess source spectra
are dominated by CO bandhead emission at 2.29$\micron$ and 2.32$\micron$.
The right panel shows a continuum-subtracted excerpt of the CO bandhead 
emission region.
The comparison between the steep decrease in CO absorption of the co-incidental
background giant neighbour (topmost black spectrum) and the slow rise in the blue wing 
of the CO emission in the excess sources provides evidence for rotational broadening.
This broadening indicates that the CO emission arises
in the inner parts of rotating circumstellar disks (Bik \& Thi 2004).
High-resolution spectroscopy will be required to quantify the rotational velocities
in the circumstellar disks.
Two of the excess sources display Br$\gamma$ in absorption, indicating that 
the extended, massive disks with strong Br$\gamma$ emission as observed towards
younger star-forming regions are already depleted in these objects.}
\end{figure*}


\begin{figure*}
\hspace*{1cm}
\includegraphics[width=12cm]{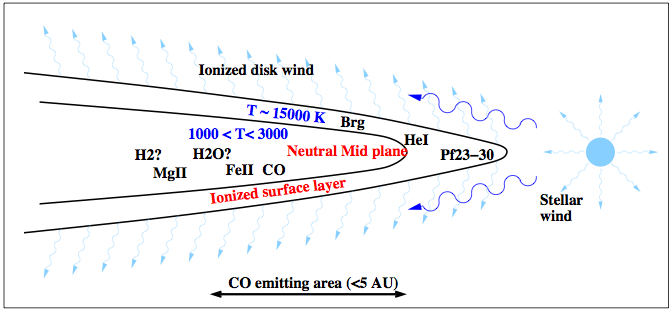}
\caption{\label{diskmodel} Schematic disk model indicating the zones 
around a massive young stellar object where CO and ionized gas emission
might originate (adopted from Bik et al.~2005). The CO bandhead emission 
originates in the dense, self-shielded region within 5 AU from the central
star, while Br$\gamma$ emission can be generated in the inner disk rim very 
close to the star as well as in the illuminated disk surface at all radii if 
the star emits sufficient UV radiation to ionize hydrogen. The lack of Br$\gamma$
emission indicates that the disk-bearing stars have spectral types later than B3V.}
\end{figure*}


\begin{figure*}
\hspace*{-1cm}
\includegraphics[width=6.4cm, angle=90]{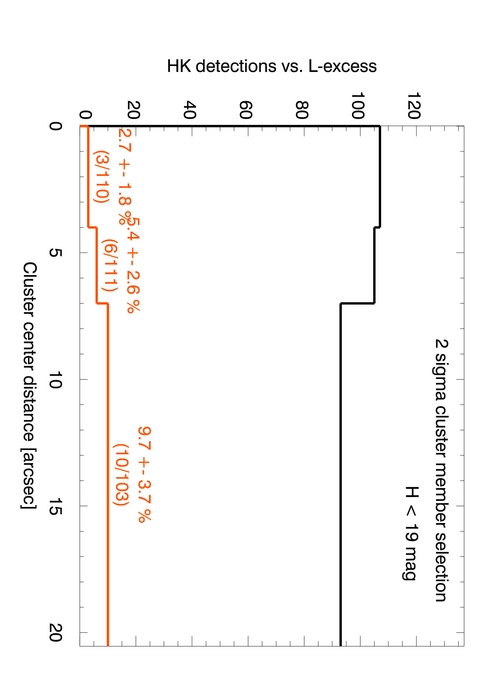}
\includegraphics[width=6.4cm, angle=90]{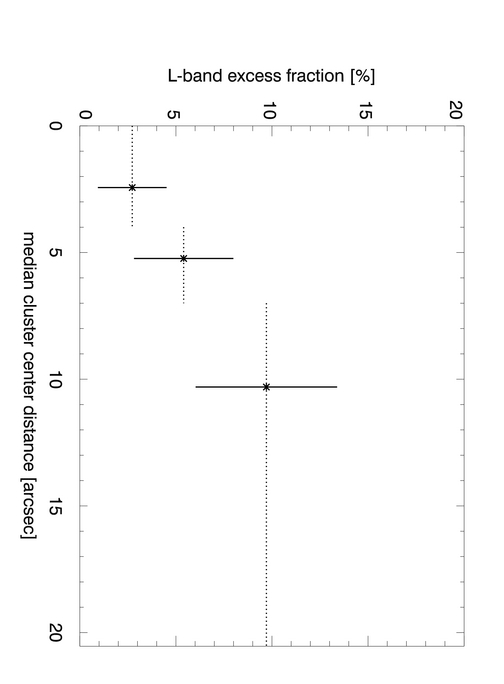}
\caption{\label{excess} Radial distribution of excess sources.
Left: Histogram of all $HK'$ detections to a limiting magnitude of $H < 19$ mag
(upper histogram) and excess sources (lower histogram). 
The number counts and relative fractions of
stars with $L'$-band excess and main sequence cluster members are labeled 
on the histogram of excess sources.
Right: $L'$-band excess fraction vs.~radial distance from the Arches cluster center.
Solid lines represent propagated uncertainties of the number counts. Dotted lines
mark the radial range covered by each bin. Bins are chosen to contain similar 
total source numbers to minimise systematic uncertainties from low number
statistics.}
\end{figure*}


\begin{figure*}
\includegraphics[width=12.4cm, angle=0]{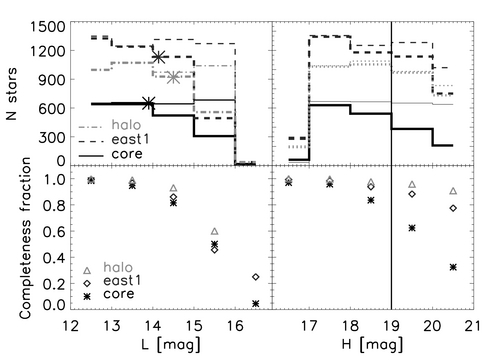}
\caption{\label{incfig} Completeness simulation results in fields core, 
east1 and halo. The observing conditions are similar for east1 and east2,
while the halo field is representative of the leading field.
Top: $L'$ and $H$ recovery number counts. Thin lines represent the number 
counts of inserted 
artificial stars in each magnitude bin, while thick lines represent the 
number counts of recovered sources. Asterisks in the $L'$ plot represent 
the faintest excess source in each field. 
Bottom: $L'$ and $H$ completeness fractions. The line at $H = 19$ mag 
shows the imposed $H$-band completeness limit
of both main sequence stars and excess sources.}
\end{figure*}


\begin{figure*}
\vspace*{-2cm}
\includegraphics[width=12.4cm]{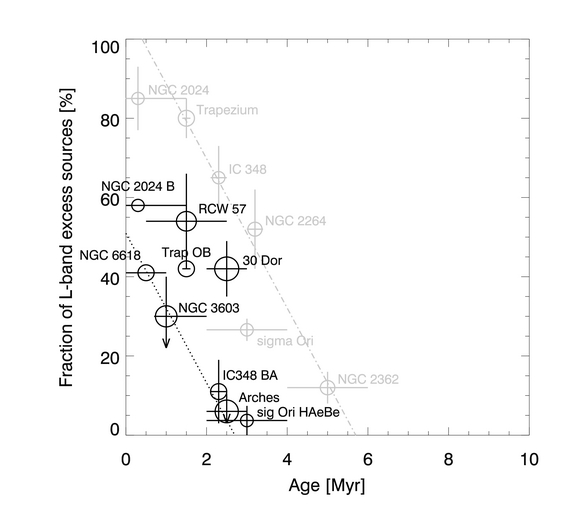} 
\caption{\label{diskage} Disk fraction vs.~cluster age reproduced from 
Haisch et al.~2001, including data points from Hern\'{a}ndez et al.~(2007) (sigma Ori), 
Hoffmeister et al.~(2006) (M17: NGC 6618), Maercker \& Burton (2005) (30 Dor region), 
and Maercker et al.~(2006) (NGC 3576: RCW 57). 
Symbols are scaled to the logarithm of the cluster mass, from the least massive
with $\sim 200\,M_\odot$ in stars (NGC\,2024, IC\,348, $\sigma$ Ori), to
the most massive with $> 30000\,M_\odot$ (30 Dor, see Table \ref{clustab}).
Black circles mark disk fractions derived from high-mass stars of types OBA
only, while light grey circles mark populations dominated by low-mass stars.
Note that the 30 Dor disk fraction covers the extended HII region, including
star-forming ridges harbouring YSO candidates, but does not resolve the central 
cluster, and is therefore an upper limit to the disk fraction in this environment.
The dash-dotted line corresponds to the linear decrease in disk fraction vs.
cluster age as fitted by Haisch et al. (lighter circles only).
In the case of the Arches and NGC\,3603, the downawrd arrow indicates 
the radial decrease in the fraction of disks from larger radii toward the cluster
core (NGC 3603 outer cluster region: Stolte et al.~2004, resolved core: 
Harayama et al.~2008). 
The dotted line is a parallel shift from the linear fit
by Haisch et al. Intriguingly, there appear to be several populations of clusters
following a similarly steep decline in disks, albeit from different initial 
disk fractions, suggesting a strong environmental effect on disk survival.
This sample is comprised of a heterogeneous set of clusters over a 
large distance and cluster mass scale, probing disks in different stellar mass 
regimes in addition to the cluster environment. The difference in the disk survival
timescales between moderate- and high-mass clusters reveals the necessity to probe
the low-mass population in massive clusters to distinguish environmental effects
from the stellar mass dependence of the disk lifetime.}
\end{figure*}


\begin{table*}
\caption{\label{fieldtab} Arches cluster fields observed with Keck/NIRC2}
\begin{tabular}{lccrrcr}
\hline
Field   & RA           & DEC          & $\delta\,$RA$^1$ & $\delta\,$DEC$^1$ & tip-tilt star$^2$ & distance to TTS\\
core    & 17:45:50.540 & -28:49:20.07 &   1.681  &  -2.190 & 425 &  10.369 \\
east1   & 17:45:51.060 & -28:49:20.07 &   8.620  &  -2.190 & 425 &  17.200 \\
east1.5$^3$ & 17:45:51.600 & -28:49:24.40 & 15.826  &   2.140 & 401 &  18.149 \\
east2   & 17:45:51.740 & -28:49:24.28 &  17.694  &   2.020 & 401 &  16.352 \\
lead    & 17:45:50.960 & -28:49:11.72 &   7.286  & -10.540 & 425 &  18.998 \\
halo    & 17:45:49.720 & -28:49:16.87 &  -9.261  &  -5.390 & 425 &   5.603 \\
\hline
\end{tabular}
$^{1}$ {\small Positional offsets in right ascension and declination are given in arcseconds, relative to 
the brigthest source in the cluster core (see Fig.~\ref{mosaic}), RA 17:45:50.42, DEC -28:49:22.3.
The cluster center is located at RA 17:45:50.54, DEC -28:49:19.8.}
$^2$ {\small Tip-tilt star identification numbers are given as extension from GSC2.2:S222122222, 
located at 425: RA = 17:45:49:78, 17:45:22.3, 401: 17:45:52.93, -28:49:28.13.}
$^3$ {\small The intermittent field east1.5 mostly overlaps with field east2, as well as a small 
section of field east1, to provide proper motion membership in the area not covered by the NACO field.}
\end{table*}


\begin{table*}
\tiny
\caption{\label{obstab} Log of Keck/NIRC2 imaging and VLT/SINFONI spectroscopy Arches observations}
\begin{tabular}{lllcrrrrcccc}
\hline
\multicolumn{11}{c}{Keck/NIRC2 imaging} \\
\hline
Date & Field & Filter & $t_{exp}$ [s] & coadds & $N_{obs}$ & $N_{used}^4$ & $t_{int}$ [s] & FWHM [mas] & Strehl$^1$ & PSF size$^3$ [pix] & LF peak$^2$ \\
\hline
2006 May 21  & core  & H  &  10 &  6 & 15 & 15 & 900 & 63.2 & 0.17 & 100 & 19.0\\
             & core  & L' & 0.5 & 60 & 15 & 15 & 450 & 79.4 & 0.64 &  47 & 14.0 \\
             & east1 & L' & 0.5 & 60 & 16 & 16 & 480 & 81.3 & 0.58 &  40 & 13.5 \\
             & GC$^5$& L' & 0.5 & 60 & 19 & 19 & 570 & 81.4 & 0.38 & 100 & 12.0 \\
\\
2006 July 18 & core  & K' &  3 & 10 & 52 & 36 & 1080 & 56.0 & 0.22 & 100 & 19.5 \\
             & east1 & H  & 10 &  6 & 16 & 14 & 840 & 67.4 & 0.10 &   80 & 20.5 \\
             & east1 & K' &  3 & 10 & 15 & 10 & 300 & 65.0 & 0.15 &   90 & 18.5 \\ 
             & east2 & H  & 20 &  3 & 15 & 13 & 780 & 71.3 & 0.08 &   70 & 20.5 \\
             & east2 & K' &  3 & 10 & 15 & 15 & 450 & 59.3 & 0.22 &   75 & 19.5 \\
             & east2 & L' & 0.5 & 60 & 17 & 17 & 510 & 79.2 & 0.69 &  40 & 14.5 \\
\\
2007 June 21 & leading   & K' & 3 & 10 & 42 & 40 & 1200 & 56.0 & 0.38 &  80 & 20.5 \\
             & halo west & K' & 3 & 10 & 28 & 26 &  780 & 53.7 & 0.38 &  80 & 19.5 \\
\\
2008 July 1  & leading   & H  &  10 &  3 & 29 & 29  &  870 & 58.7 & 0.20 & 100 & 20.5 \\
             & leading   & L' & 0.5 & 60 & 53 & 53  & 1590 & 80.8 & 0.55 &  80 & 15.5 \\
             & halo west & H  & 10  &  3 & 23 & 23  &  690 & 50.5 & 0.23 & 100 & 20.5 \\
             & halo west & L' & 0.5 & 60 & 39 & 39  & 1170 & 78.8 & 0.76 &  80 & 14.5 \\
             & east1.5   & K' &  3  & 10 & 22 & 21  &  630 & 68.0 & 0.29 & 100 & 19.5 \\
\hline
\multicolumn{11}{c}{VLT/SINFONI integral-field spectroscopy} \\
\hline
2006 May-July & core & HK & 300 & 24 & - & 7200 &  -  & 135 & - & - & - \\
\hline
\end{tabular}
\footnotesize{$^1$ FWHM and Strehl values are estimated from the averaged PSF
image derived for each frame.}
\footnotesize{$^2$ The peak of the luminosity function indicates the 
completeness in each filter on each field.}
\footnotesize{$^3$ The PSF size is the diameter of the PSF extracted with starfinder and used to perform PSF fitting across each field (see Sec.~\ref{photsec}).}
\footnotesize{$^4$ Frames for the combined image were selected on the basis of the FWHM and the
AO performance. For observing sequences with less than 20 frames, all frames with AO correction 
are included in the final image, while for sequences with more than 20 frames selection based on 
the FWHM in each image was possible (see Sec.~\ref{obssec} for details).}
{$^5$ GC refers to the Galactic center data set used for $L'$ calibration.}
\end{table*}


\begin{table*}
\caption{\label{zpttab} Residual zeropoint uncertainties}
\begin{tabular}{lcrcrcrl}
\hline
field & $\sigma_H$ & $N_H^1$ & $\sigma_{K'}$ & $N_{K'}^1$ & $\sigma_{L'}$ & $N_{L'}^1$  & calibration field \\
      &  [mag]     &         &   [mag]    &         &  [mag]     &          &                   \\
\hline
core$^2$    & 0.010 &  56 & 0.009 &  64 & 0.100 & -  & $HK'$ NACO/NIRC2 core \\
            &       &     &       &     &       &    & $L'$ Galactic center zeropoint \\
east1       & 0.012 & 100 & 0.012 &  72 & 0.100 & 31 & $HK'L'$ core \\ 
east1.5     &    -  &  -  & 0.022 &  22 &   -   &    & $K'$ core \\
east2       & 0.024 &  17 & 0.023 & 103 & 0.101 &  6 & $HL'$ core, $K'$ east1.5 \\
lead        & 0.019 &  29 & 0.017 &  29 & 0.107 &  7 & $HK'L'$ core \\
halo        & 0.010 & 107 & 0.006 & 228 & 0.101 & 30 & $HK'$ NACO, $L'$ core \\
\hline
\end{tabular}
\\
$^1$ Number of calibration sources available. In the core $L'$ data, 
the Galactic center zeropoint was applied (see Appendix B). 
$^2$ The core $HK'$ zeropoint was derived from the NACO central region, $250 < x,y < 750$ pixels (NIRC2) 
to minimise anisoplanatic and -kinetic effects. The core $L'$ zeropoint was derived from the 
Galactic center standard field, all adjacent fields were calibrated with respect to the core field. 
The standard deviation in photometric residuals after calibration is given as the $HK'$ zeropoint 
uncertainties. The $L'$ uncertainty is the zeropoint variation during the GC observing sequence in 
the core field, and the variation between calibration sources is taken into account for the surrounding fields.
\end{table*}


\begin{table*}
\caption{\label{hkltab} L-band excess sources in the Arches cluster}
\begin{tabular}{lccrrrrrrcc}
\hline
ID & $\delta\,$RA & $\delta\,$DEC$^{1}$ & $H$ & $\sigma_H$ & $K'$ & $\sigma_K$ & $L'$ & $\sigma_L$ & $A_V^2$ & Spectra \\
   & ['']       &   ['']            & [mag] &  [mag]   & [mag]& [mag]      & [mag]&  [mag]     & [mag]   &          \\
\hline
1 &   5.852 &    2.239 &  17.220 &  0.059 & 15.040 &  0.011 & 12.205 &  0.103 & $25.4 \pm 2.1$ & \\
2 &   3.328 &    2.113 &  17.227 &  0.014 & 15.192 &  0.009 & 12.514 &  0.100 & $25.8 \pm 0.5$ & Sp \\
3 & -2.948 &    3.563 &  17.848 &  0.064 & 15.458 &  0.010 & 12.486 &  0.103 & $27.4 \pm 0.6$ & Sp \\
4 &  3.947 &    1.106 &  17.979 &  0.025 & 15.915 &  0.010 & 13.167 &  0.100 & $25.8 \pm 0.7$ & Sp \\
5 & -0.963 &    0.526 &  18.071 &  0.019 & 16.030 &  0.011 & 13.672 &  0.101 & $25.1 \pm 0.7$ &  \\
6 & -2.993 &   -0.440 &  18.656 &  0.178 & 16.456 &  0.058 & 13.878 &  0.103 & $25.1 \pm 1.3$ &  \\
7 &  7.064 &   -0.835 &  17.521 &  0.022 & 15.555 &  0.018 & 12.828 &  0.104 & $25.5 \pm 1.3$ & \\
8 & 11.637 &   -2.263 &  17.962 &  0.022 & 16.087 &  0.020 & 13.488 &  0.103 & $23.3 \pm 0.5$ &  \\
9 &  7.470 &    2.160 &  18.502 &  0.014 & 16.397 &  0.024 & 13.728 &  0.104 & $24.5 \pm 0.7$ & \\
10 &  7.850 &    1.769 &  18.095 &  0.020 & 16.215 &  0.016 & 14.112 &  0.102 & $25.5 \pm 1.6$ &  \\
11 & 11.772 &   -0.970 &  18.691 &  0.023 & 16.585 &  0.017 & 13.622 &  0.113 & $23.4 \pm 0.7$ &  \\
12 & 14.877 &   -1.496 &  17.459 &  0.038 & 15.526 &  0.023 & 13.045 &  0.102 & $23.6 \pm 0.6$ &  \\
13 & 15.480 &   -4.183 &  17.630 &  0.026 & 15.630 &  0.023 & 12.982 &  0.102 & $23.7 \pm 1.0$ & \\
14 &  5.329 &    7.119 &  16.659 &  0.032 & 14.654 &  0.024 & 11.893 &  0.107 & $25.7 \pm 1.2$ &  \\   
15 &  3.512 &   15.159 &  17.303 &  0.031 & 15.069 &  0.032 & 12.056 &  0.110 & $25.2 \pm 2.6$ &  \\
16 &  8.788 &    7.212 &  17.487 &  0.029 & 15.443 &  0.017 & 12.927 &  0.107 & $24.8 \pm 1.6$ &  \\
17 &  9.011 &   13.247 &  18.987 &  0.020 & 16.509 &  0.026 & 13.133 &  0.108 & $26.2 \pm 1.9$ &  \\
18 &  2.435 &   14.149 &  18.965 &  0.035 & 17.128 &  0.029 & 14.593 &  0.132 & $24.7 \pm 1.5$ &  \\
19 &-12.619 &    1.492 &  18.177 &  0.026 & 15.529 &  0.013 & 12.293 &  0.101 & $24.8 \pm 2.4$ &  \\
20 & -5.101 &    4.691 &  18.318 &  0.013 & 15.643 &  0.020 & 12.366 &  0.103 & $27.9 \pm 0.6$ &  \\
21 & -4.151 &    4.383 &  19.958 &  0.018 & 17.373 &  0.044 & 14.351 &  0.102 & $27.7 \pm 0.8$ &  \\
22 &-11.824 &    2.234 &  20.676 &  0.021 & 18.123 &  0.009 & 15.317 &  0.116 & $24.8 \pm 2.4$ & \\
23 & 16.215 &    0.249 &  18.690 &  0.025 & 16.907 &  0.025 & 14.875 &  0.111 & $22.7 \pm 0.6$ & \\  
24 &  4.240  &   2.615 &  17.366 &  0.051 & 15.607 &  0.012 & 13.527 &  0.104 & $26.3 \pm 1.0$ & \\
\hline
\end{tabular}
\\
$^{1}$ {\small Positional offsets in right ascension and declination are given in arcseconds, relative to 
the brigthest source in the cluster core (see Fig.~\ref{mosaic}), RA 17:45:50.42, DEC -28:49:22.3.
The cluster center is located at RA 17:45:50.54, DEC -28:49:19.8.}
\\
$^{2}$ {\small The visual extinction is estimated as an average of the extinction of the 4 nearest
main sequence ($1.4 < H-K' < 1.9$ mag) 
cluster members using a Rieke \& Lebofsky 1985 extinction law.}
\end{table*}


\begin{table*}
\caption{\label{wrtab} L-band properties of Wolf-Rayet stars}
\begin{tabular}{lccrrrrrrlc}
\hline
ID & $\delta\,$RA & $\delta\,$DEC$^{1}$ & $H$ & $\sigma_H$ & $K'$ & $\sigma_K$ & $L'$ & $\sigma_L$ & SpT & Ref.$^2$ \\
   & ['']       &   ['']            & [mag] &  [mag]   & [mag]& [mag]      & [mag]&  [mag]     &     &      \\
\hline
1 &  0.000 &    0.000 &  12.023 &  0.030 & 10.258 &  0.179 & 8.364 &  0.100 & WN8-9h & 1 \\
2 &  1.936 &    4.725 &  12.097 &  0.044 & 10.447 &  0.090 & 8.577 &  0.101 & WN8-9h & 1 \\
3 &  0.664 &    2.764 &  12.145 &  0.021 & 10.535 &  0.079 & 8.749 &  0.100 & WN8-9h & 1 \\
4 & -0.419 &    1.048 &  12.465 &  0.013 & 10.772 &  0.021 & 8.822 &  0.100 & WN8-9h & 1 \\
5 &  3.381 &   -0.302 &  12.819 &  0.014 & 11.144 &  0.011 & 9.443 &  0.100 & WN7/OIf* & 2,3 \\
6 &  1.348 &    1.620 &  12.925 &  0.013 & 11.349 &  0.014 & 9.763 &  0.100 & O4-6If* & 1 \\
7 & -1.893 &    5.057 &  12.992 &  0.025 & 11.185 &  0.025 & 9.414 &  0.100 & WN7 & 2 \\
8 &  4.522 &    8.055 &  12.390 &  0.023 & 10.497 &  0.030 & 8.530 &  0.108 &  - & - \\
\hline
\end{tabular}
\\
$^{1}$ {\small Positional offsets in right ascension and declination are given in arcseconds, relative to 
the brigthest source in the cluster core (see Fig.~\ref{mosaic}), RA 17:45:50.42, DEC -28:49:22.3.
The cluster center is located at RA 17:45:50.54, DEC -28:49:19.8.}
$^{2}$ {\small References for spectral types: 1 - Martins et al.~2008; 2 - Blum et al.~2001; 3 - Figer et al.~2002}
\end{table*}


\begin{landscape}
\begin{table}
\small
\caption{\label{clustab} Physical properties and disk fractions of young star clusters}
\begin{tabular}{llcccccccccl}
\hline
Name & location & dist & age & $M_{cl}$ & $r_{core}$ & $\rho_{core}$ & \multicolumn{2}{c}{disk fraction} &  mass range & Ref \\
     &          & kpc  & Myr & $M_\odot$ & pc & $M_\odot {\rm pc^{-3}}$ & all & OB(A) stars & $M_\odot$ &  \\
\hline
\multicolumn{7}{c}{Starburst clusters $M_{cl} > 10^4\,M_\odot$} \\
\hline
Arches     & GC &   8000            & $2.5\pm 0.5$   & $>2\cdot 10^4$ & $0.14\pm 0.05$ & $2\cdot 10^5$   & - & $6 \pm 2$ & $3 - 20\,M_\odot$ & 1,2,3,4 \\
Quintuplet & GC &   8            & $4\pm 1$       & $2\cdot 10^4$ &  -    &      -      & - & - & - &  5 \\
NGC\,3603\,YC& SP &   6-7        & 1              & $10^4$   & 0.2   & $\sim10^5$ & $27\pm 3$ & $22\pm 10$ & $1.2-20\,M_\odot$ & 6,7 \\
Westerlund 1 & SP & $3.5\pm 0.2$ & $4\pm 1$       &  52000    & $< 1$ & $>2\cdot 10^4$  & - & - & - & 8 \\
R\,136/30 Dor& LMC& 50           & 2-3            & $>3\cdot 10^4$ & 0.24  & $\sim 10^5$   & - & $42\pm 5$ & $> 20\,M_\odot$ & 9,10 \\
\hline
\multicolumn{7}{c}{Young clusters $M_{cl} < 10^4\,M_\odot$} \\
\hline
ONC/Trap         & SP & 0.43  & 0.3-1 & $10^3$ & 0.2 & $4\ 10^4$      & 80 & 42   & $0.2-35\,M_\odot$ & 11,12 \\
NGC\,2024        & SP & 0.46  & 0.3-1 & 200  & -     & $6\cdot 10^3$  & $85\pm 5$ & 58 & $0.1-20\,M_\odot$ & 12 \\
$\sigma$ Ori     & SP & 0.5   & 2.5-3 & 225  & 1.6   & 4              & $27 \pm 3$ & $4 \pm 4$ & $0.1-3\,M_\odot$ & 13, 14 \\ 
NGC\,3576/RCW 57 & SP & 2.8   & $1.5\pm 1$ & $5\cdot 10^3$ & - & $3\cdot 10^3$ & -  & $55\pm 12$ & $10-35\,M_\odot$ & 15, 16 \\
M 17/NGC\,6618   & SP & 2.1   & 0.3-1 & $>10^3$ & -  & $> 60$ & 62 & 41 & $2-90\,M_\odot$ & 17 \\
NGC\,2264        & SP & 0.8   & 3.2   & $> 430$ & -  &  -     & $52\pm 10$ & - & $0.8-30\,M_\odot$ & 12, 18 \\
IC\,348          & SP & 0.3   & 2-3  & 208  & 0.1 & $2\cdot 10^3$ & $30\pm 4$ & $< 11\pm 8$ & $0.2-6\,M_\odot$ & 19, 20 \\
NGC\,2362        & SP & 1.5  &  5    & 500  & - & - & $7\pm 2$ & 0.0  & $0.3-10\,M_\odot$ & 21, 22 \\ 
\hline
\end{tabular}
1 - Espinoza et al.~2009; 2 - Stolte et al.~2005; 3 - Najarro et al.~2004; 4 - Figer et al.~1999a;
5 - Figer et al.~1999b; 6 - Harayama et al.~2008; 7 - Stolte et al.~2004; 
8 - Brandner et al.~2008; 9 - Brandl et al.~1996; 10 - Maercker \& Burton 2005; 
11 - Hillenbrand \& Hartmann (1998); 12 - Haisch et al.~(2000,2001); 
13 - Hern\'{a}ndez et al.~2007; 14 - Sherry et al.~2004; 
15 - Figuer\^{e}do et al.~2002; 16 - Maercker et al.~2006; 
17 - Hoffmeister et al.~2006; 18 - Dahm \& Simon 2005; 
19 - Lada et al.~2006; 20 - Lada \& Lada 1995; 
21 - Dahm 2005; 22 - Dahm \& Hillenbrand 2007
\end{table}
\end{landscape}

\clearpage


\begin{deluxetable}{llrrrrrrrrrrrrrcc}
\rotate
\tabletypesize{\scriptsize}
\tablewidth{0pt}
\tablecolumns{17}
\tablecaption{\label{fulltable} Photometry and astrometry source list of the Arches cluster.}
\tablehead{%
\colhead{ID} & 
\colhead{field} & 
\colhead{$\delta\,$RA} & 
\colhead{$\delta\,$DEC} & 
\colhead{$\sigma_x$} & 
\colhead{$\sigma_y$} & 
\colhead{$H$} & 
\colhead{$\sigma_H$} & 
\colhead{$K_s$} & 
\colhead{$\sigma_{K_s}$} & 
\colhead{$L'$} & 
\colhead{$\sigma_{L'}$} & 
\colhead{baseline} & 
\colhead{$\mu_{\alpha cos(\delta)}$} & 
\colhead{$\mu_\delta$} & 
\colhead{member} & 
\colhead{excess} \\
\colhead{ } & \colhead{ } & \colhead{[ '' ]}  & \colhead{[ '' ]} & \colhead{[ '' ]} & \colhead{[ '' ]} & \colhead{mag} & \colhead{mag} & \colhead{mag} & \colhead{mag} & \colhead{mag} & \colhead{mag} & \colhead{years} & \colhead{mas/yr} & \colhead{mas/yr} & \colhead{ } & \colhead{ }}
\startdata
1  &    core    &     0.000   &   0.000   &  0.0010   &  0.0001  &  12.023  & 0.030  & 10.258  & 0.179   &   8.364   &   0.100  &  4.33   &  -0.551   &   0.046   &   1    &   0 \\
2  &    core    &     1.936   &   4.725   &  0.0013   &  0.0013  &  12.097  & 0.044  & 10.447  & 0.090   &   8.577   &   0.101  &  4.33   &  -0.597   &  -0.115   &   1    &   0 \\
3  &    core    &     0.664   &   2.764   &  0.0004   &  0.0004  &  12.145  & 0.021  & 10.535  & 0.079   &   8.749   &   0.100  &  4.33   &  -0.069   &   0.023   &   1    &   0 \\
4  &    core    &    -0.419   &   1.048   &  0.0003   &  0.0002  &  12.465  & 0.013  & 10.772  & 0.021   &   8.822   &   0.100  &  4.33   &   0.000   &   0.345   &   1    &   0 \\
5  &    core    &     3.381   &  -0.302   &  0.0003   &  0.0001  &  12.819  & 0.014  & 11.144  & 0.011   &   9.443   &   0.100  &  4.33   &  -0.092   &  -0.184   &   1    &   0 \\
6  &    core    &     1.348   &   1.620   &  0.0005   &  0.0002  &  12.925  & 0.013  & 11.349  & 0.014   &   9.763   &   0.100  &  4.33   &  -0.368   &  -0.161   &   1    &   0 \\
7  &    core    &    -1.893   &   5.057   &  0.0003   &  0.0003  &  12.992  & 0.025  & 11.185  & 0.025   &   9.414   &   0.100  &  4.33   &   0.092   &   0.368   &   1    &   0 \\
\enddata
\newline
\tablecomments{The full table is available in the online version of the journal.}
\end{deluxetable}

\end{document}